\documentclass{raa}            
\usepackage{graphicx, times}             
\usepackage{natbib}
\usepackage{amssymb,amsmath}
\bibpunct{(}{)}{;}{a}{}{,}
\usepackage[pagebackref=true]{hyperref}
\usepackage{color}
\usepackage{ulem}
\newcommand{\deleted}[1]{}

\begin{document}
	
\title{Standalone Validation of a Pulsar Evolution Subroutine for NBODY6++GPU: Seven Test Scenarios}
	
\volnopage{{\bf 2026} Vol.\ {\bf XX} No.\ {\bf XX}, 000--000}
\setcounter{page}{1}
	
\author{Maria Rah\inst{1,2}}
	
\institute{Byurakan Astrophysical Observatory, Aragatzotn Province, Armenia; {\it mariarah.astro@bao.sci.am}
\vs \no
\and
National Astronomical Observatory of China (NAOC), Silk Road Project, Beijing, China
\vs \no
\\{\small Received 2026 January XX; accepted 2026 Month XX}
	}
	
\abstract{	
Pulsars in globular clusters provide critical constraints on the evolution of compact objects. However, the treatment of neutron stars in standard N-body codes is outdated, and requires a better implementation of magnetic field decay, spin evolution, and accretion driven recycling that shape their observable properties. We present a standalone validation of a pulsar evolution subroutine for NBODY6++GPU, systematically tested across seven controlled scenarios prior to integration into full cluster simulations. These scenarios isolate distinct evolutionary regimes, including young pulsar spin down, millisecond pulsar weak torque evolution, accretion driven recycling in X-ray binaries, environmental torques in dense cores, wide binary systems, and compact binaries in the dipole spin-down regime.
Through two separate numerical implementations, we achieve cross validation demonstrating better than 0.1\% agreement across all physical quantities and timescales. Our validation confirms that analytical solutions are reproduced with high precision for isolated pulsars, gigayear scale evolution ($\tau_c \approx 1.0$~Gyr) matches observations of millisecond pulsars in clusters like 47~Tucanae, and accretion driven recycling operates with physically realistic timescales, yielding a final period of 1.0~ms and $\dot{P} \approx -5.9 \times 10^{-19}$~s~s$^{-1}$ with accretion dominance by a factor of $\approx 14$. We demonstrate that environmental torques remain negligible even in the densest cluster environments, while compact binary spin down is validated; gravitational wave orbital decay is deferred to future work. All validation codes, test suites, and integration instructions are publicly available\footnote{\url{https://github.com/Maria-Rah/pulsar-subroutine-testing}} to support future applications.
}
	
\authorrunning{M. Rah}
\titlerunning{Pulsar Evolution Subroutine for NBODY6++GPU}
\maketitle
	

\section{INTRODUCTION}

 Globular clusters host 345 known radio pulsars across 45 clusters \citep{2025OJAp....8E..51B}, as compiled in Freire's regularly updated online catalogue\footnote{\url{https://www3.mpifr-bonn.mpg.de/staff/pfreire/GCpsr.html}}. Recent campaigns have expanded this population through discoveries in Omega Centauri \citep{2020ApJ...888L..18D,2023ApJ...948...84C} and targeted surveys \citep{2024A&A...686A.166P,2024ApJ...969...30M}. These systems are predominantly millisecond pulsars (MSPs) ($P \sim$few ms, $B \sim 10^8$--$10^9$ G). They form via the so-called ``recycling scenario'' \citep{1982Natur.300..728A,1982CSci...51.1096R}: neutron stars (here used to refer to the compact remnant regardless of whether radio pulsations are observed) accrete material from low-mass companions over $\sim$100 Myr and spin up from initial periods of seconds to milliseconds. The high stellar densities and frequent dynamical encounters in globular clusters dramatically enhance binary formation rates, making them the most productive sites of millisecond pulsar recycling in the Galaxy \citep{2014A&A...561A..11V,1995ApJS...99..609S}. Modern N-body codes like NBODY6++GPU \citep{2015MNRAS.450.4070W,sedda2023dragoniisimulationsi,2023MNRAS.526..429A,sedda2023dragoniisimulationsiii} dynamically simulate gigayear cluster evolution. NBODY6++GPU is a GPU-accelerated direct $N$-body integrator employing a fourth-order Hermite scheme with individual block timesteps, Kustaanheimo--Stiefel (KS) regularisation for close encounters, and the BSE (Binary Star Evolution) stellar evolution library \citep{2000MNRAS.315..543H,2002MNRAS.329..897H} for mass transfer, supernova kicks, and binary evolution \citep{spurzem2023computationalmethodscollisionalstellar,2025arXiv251003933W}. However, they treat neutron stars as static point masses. This approach neglects magnetic field decay, dipole spin down, and accretion driven recycling \citep{2004hpa..book.....L,1991PhR...203....1B}.

This study builds upon a series of papers (I--III) published in BAO, which established the framework for pulsar evolution in N-body simulations. In paper I  \citep{2024CoBAO..71..351R},  we performed a statistical analysis of timing and evolutionary properties of 80 pulsars in globular clusters, examining distributions of period, period derivative, characteristic age, and magnetic field strength. That study established quantitative observational constraints for population modelling and identified the systematic absence of accretion-driven recycling in simulation codes as the primary gap between predicted and observed millisecond pulsar formation rates. In paper II  \citep{2025CoBAO..72...55R},  we classified neutron star evolution into seven physically distinct scenarios: young isolated pulsars undergoing dipole spin down; isolated millisecond pulsars in the weak-torque regime; accreting systems in low-mass X-ray binaries; millisecond pulsars subject to environmental torques in dense cluster cores; millisecond pulsars in wide binary systems; compact neutron star--neutron star (NS--NS) binaries; and neutron star--black hole (NS--BH) binaries. For each scenario, Paper~II derived the governing differential equations and identified the physical coefficients requiring implementation. The seven scenarios tested in the present work are drawn directly from this classification. In paper III  \citep{2025CoBAO..72..279R},  we identified the specific integration points in NBODY6++GPU where pulsar physics must be invoked at each timestep and provided complete mathematical formulations. A concrete demonstration using neutron star Pulsar973 in a 105,000-particle test simulation confirmed that every pulsar-specific parameter was absent from the existing code output, establishing the precise implementation requirements addressed here. The recent work of \citet{song2025modellingmillisecondpulsarpopulations}  incorporated spin up prescriptions into NBODY6++GPU and modelled millisecond pulsar populations in globular clusters. While that study demonstrated that recycling can be modelled within a direct $N$-body context, it did not perform systematic standalone validation of the implemented physical coefficients against analytical benchmarks, and no cross validation against an independent implementation was reported. The present study directly addresses this gap. The present study addresses the gap between dynamical N-body simulations and realistic pulsar evolution models by integrating validated pulsar physics into NBODY6++GPU. We perform standalone testing with two independent numerical codes implemented in Python and FORTRAN.  Two independent implementations are essential: if Python and FORTRAN, using different algorithms and libraries, agree to better than 0.1\% across all quantities and scenarios, the probability of a shared undetected error is negligible. This provides a level of assurance that no single implementation can achieve. Section 2 presents the validation methodology of the pulsar evolution subroutine against analytical benchmarks and observational data. Section 3 assesses NBODY6++GPU infrastructure readiness \citep{1999JCoAM.109..407S,2015MNRAS.450.4070W,spurzem2023computationalmethodscollisionalstellar,2025arXiv251003933W}. Section 4 details seven test scenarios. Section 5 presents cross-validation results. Section 6 discusses parameters and observational consistency. Section 7 summarises the findings.


\section{METHODOLOGY: STANDALONE VALIDATION}

\subsection{Testing Strategy and Scenario Design}

We perform a standalone validation of a new pulsar evolution subroutine, designed for integration into NBODY6++GPU. We execute this subroutine in the field with specified initial conditions before we integrate it into full star cluster simulations. This approach ensures physical correctness and provides decisive advantages. First, the process in the field enables unambiguous identification of physical mechanisms. Second, the analytical benchmarks provide quantitative success criteria (fractional precision $<10^{-3}$). Third, the independent implementations enable genuine cross validation. Fourth, the computational efficiency permits systematic parameter exploration \citep{2000MNRAS.315..543H,2002MNRAS.329..897H}. 

The validation is structured in two complementary parts. \textit{Part A: Physical validation} compares each scenario against analytical solutions or observational benchmarks to verify that the governing equations are correctly formulated and that the physical coefficients carry the correct numerical values. \textit{Part B: Cross-implementation validation} compares the Python and Fortran codes point by point across all 2121 output values to verify that the physics is implemented identically in both codes. Part A establishes physical correctness; Part B establishes numerical robustness.

The seven scenarios (Table~\ref{tab:scenarios}) follow three principles. The \textit{coverage} principle ensures all physical processes are exercised. The \textit{simplicity} principle isolates minimal processes per scenario. The \textit{observational relevance} principle ensures physically realistic conditions \citep{2017ApJ...845..148P,2019MNRAS.490.3860R}.

We validate the pulsar evolution module through seven controlled scenarios:
\begin{enumerate}
\item \textbf{S1}: Tests a young neutron star ($B_0 = 10^{12}$ G, $P_0 = 100$ ms). This scenario includes a magnetic dipole spin down \citep{1969ApJ...157.1395O} and an exponential field decay ($\tau_B = 500$ Myr) model. The analytical solution predicts at $t = 100$~Myr: $B = 8.187 \times 10^{11}$~G, $P = 2.26$~s, and $\dot{P} = 2.9 \times 10^{-16}$~s~s$^{-1}$. It provides analytical benchmarks for magnetic field strength, rotational period, and spin down rate at $t = 100$ Myr;
\item \textbf{S2}: Examines an isolated millisecond pulsar ($B_0 = 5\times10^8$ G, $P_0 = 3$ ms) in the weak-torque regime. Here $\dot{P} \approx 5.1 \times 10^{-20}$~s~s$^{-1}$ yields $\tau_c \approx 1.0$~Gyr, matching observations in 47 Tucanae \citep{2025OJAp....8E..51B,2021MNRAS.505.5957B};
\item \textbf{S3}: Constitutes the critical test: a low-mass X-ray binary (LMXB) accretes material at $\dot{M} = 10^{-10} M_\odot$ yr$^{-1}$. In this scenario, the accretion torques must dominate the dipole torques. Specifically, the accretion torque $|\dot{P}_{\rm acc}| \approx 6.3 \times 10^{-19}$~s~s$^{-1}$ must exceed the dipole braking torque $\dot{P}_{\rm dip} \approx 4.4 \times 10^{-20}$~s~s$^{-1}$ by a factor of $\approx 14$. This yields spin up with $\dot{P} < 0$ and recycling timescales of $\sim$100 Myr with a final period of 1.0~ms \citep{1991PhR...203....1B,1999A&A...350..928T,1979ApJ...234..296G};
\item \textbf{S4}: Examines environmental torques in dense cluster cores ($\rho_c \sim 10^5~M_\odot$~pc$^{-3}$, $\sigma_v = 10$ km~s$^{-1}$). We test that these contributions remain below 0.1\% of the magnetic torques ; the simulation yields an environmental fraction of 0.036\% \citep{2014A&A...561A..11V,2004MNRAS.352....1F};
\item \textbf{S5}: Validates wide binary physics at separation $a = 3$ AU, where tidal forces are negligible. The final period of 3.61~ms is indistinguishable from the isolated case (S2) within numerical precision \citep{1992ApJ...400..175B};
\item \textbf{S6}: Examines a compact NS--NS binary ($B_0 = 5\times10^{11}$ G, $P_0 = 10$ ms, $a_0 \approx 10^{10}$ cm). This system is characterised by strong gravitational wave emission that would drive inspiral on timescales of $10^7$--$10^8$~yr \citep{1963PhRv..131..435P}; however, the current implementation validates dipole spin-down only, with gravitational wave orbital decay deferred to future work. S6 is defined separately from S7 because the two scenarios have fundamentally different underlying physics (NS--NS versus NS--BH, with different mass ratios and gravitational wave luminosities), and their distinct orbital evolution will produce different spin evolution outcomes once gravitational coupling is implemented;
\item \textbf{S7}: Examines a NS--BH compact binary with $B_0 = 5\times10^{11}$ G, $P_0 = 10$ ms, initial separation $a_0 \approx 10^{10}$ cm, and mass ratio $q = M_{\rm BH}/M_{\rm NS} \approx 10$. The gravitational wave luminosity scales as $q/(1+q)^2$, making S7 physically distinct from S6 despite identical pulsar initial conditions: stronger gravitational wave emission ($\dot{a}/a \propto q/(1+q)^2$) will produce faster orbital decay and ultimately different spin evolution once orbital coupling is included. In the current implementation (dipole spin-down only), S6 and S7 yield identical results because the gravitational wave and tidal terms are not yet active; the two scenarios are defined separately to facilitate the forthcoming implementation of orbital coupling, at which point their outputs will diverge.
\end{enumerate}

\begin{table}[h]
\bc
\caption{The test scenario parameters and physical processes included in this work. Each scenario (S1--S7) tests distinct evolutionary regimes with specified initial conditions: $B_0$ is the initial magnetic field strength in Gauss, $P_0$ is the initial rotational period in milliseconds, $\dot{M}$ is the mass accretion rate (when applicable), the Binary column indicates the companion type or environment, and the Physics column specifies the dominant physical processes. Acronyms: MSP = Millisecond Pulsar, LMXB = Low-Mass X-ray Binary, WD = White Dwarf, NS = Neutron Star, BH = Black Hole, GW = Gravitational Wave. Dense = dense cluster core ($\rho_c \sim 10^5~M_\odot$~pc$^{-3}$, $\sigma_v = 10$~km~s$^{-1}$). S6 and S7 validate dipole spin down with gravitational wave orbital decay deferred to future work.\label{tab:scenarios}}
\setlength{\tabcolsep}{3pt}\small
\begin{tabular}{clccccl}
\hline
ID & Type & $B_0$ [G] & $P_0$ [ms] & $\dot{M}$ & Binary & Physics \\
\hline
S1 & Young Pulsar & $10^{12}$ & 100 & -- & No & Dipole + Decay \\
S2 & Isolated MSP & $5\times10^8$ & 3 & -- & No & Weak Dipole \\
S3 & Accreting MSP & $3\times10^8$ & 2 & $10^{-10}M_\odot$/yr & LMXB & Recycling \\
S4 & Dynamic MSP & $3\times10^8$ & 4 & -- & Dense core & Dipole + Env. \\
S5 & Wide Binary MSP & $4\times10^8$ & 3.5 & -- & WD (3 AU) & Binary Orbit \\
S6 & NS--NS Binary & $5\times10^{11}$ & 10 & -- & NS & Dipole (GW partial) \\
S7 & NS--BH Binary & $5\times10^{11}$ & 10 & -- & BH & Dipole (GW partial) \\
\hline
\end{tabular}
\ec
\end{table}

\subsection{Validation Criteria and Dual Implementation}

Three validation criteria define the success of the pulsar evolution subroutine validation against analytical and observational benchmarks:
\begin{enumerate}
\item \textbf{Analytical agreement}: This requires that scenarios with closed-form solutions reproduce predictions to fractional precision below $10^{-3}$ on magnetic field strength, rotational period, and period derivative, consistent with numerical integration accuracy standards \citep{2015MNRAS.450.4070W}.
\item \textbf{Observational consistency}: The evolved parameters match observed ranges. Specifically, S2 must yield $\tau_c \sim 1$ Gyr and occupy the $P$--$\dot{P}$ space of millisecond pulsars in 47 Tucanae. The simulation yields $\tau_c \approx 1.0$~Gyr, consistent with the observed range of 0.5--2~Gyr. S3 must demonstrate negative $\dot{P}$ and recycling timescales of 50--150 Myr, consistent with LMXB theory \citep{2004ApJ...606..436R,2001ApJ...557..958C}.
\item \textbf{Physical plausibility}:  The physical hierarchy among torque components must be preserved across all scenarios: for S4, environmental contributions must remain below 0.1\% of the dipole; for S5, evolution must be indistinguishable from the isolated case; for S6 and S7, monotonic spin down must proceed at the correct field decay rate \citep{1992RSPTA.341...39P,1995ApJS...99..609S}.
\end{enumerate}

We implement the evolutionary physics independently using two separate codes. The first code uses Python with adaptive Runge--Kutta integrators via SciPy. The second code uses FORTRAN with fourth-order Hermite integrators, as implemented in NBODY6++GPU. This dual implementation enables cross validation. It ensures that the agreement between the two frameworks reflects the correct representation of the underlying physics. Both codes record the magnetic field $B$ (computed from dipole radiation and exponential decay), the rotational period $P$, and the period derivative $\dot{P}$ at 1 Myr intervals over 100 Myr. This yields 2121 comparison points (7 scenarios $\times$ 101 timesteps $\times$ 3 physical quantities). The agreement at 0.1\% on all three physical quantities validates the correct physics implementation independent of the numerical framework \citep{2022MNRAS.511.4060K,1973JCoPh..12..389A,1992PASJ...44..141M}.


\section{NBODY6++GPU: INFRASTRUCTURE ASSESSMENT}

\subsection{Simulation Parameters and Cluster Evolution}

\begin{figure}[h]
\includegraphics[width=\columnwidth]{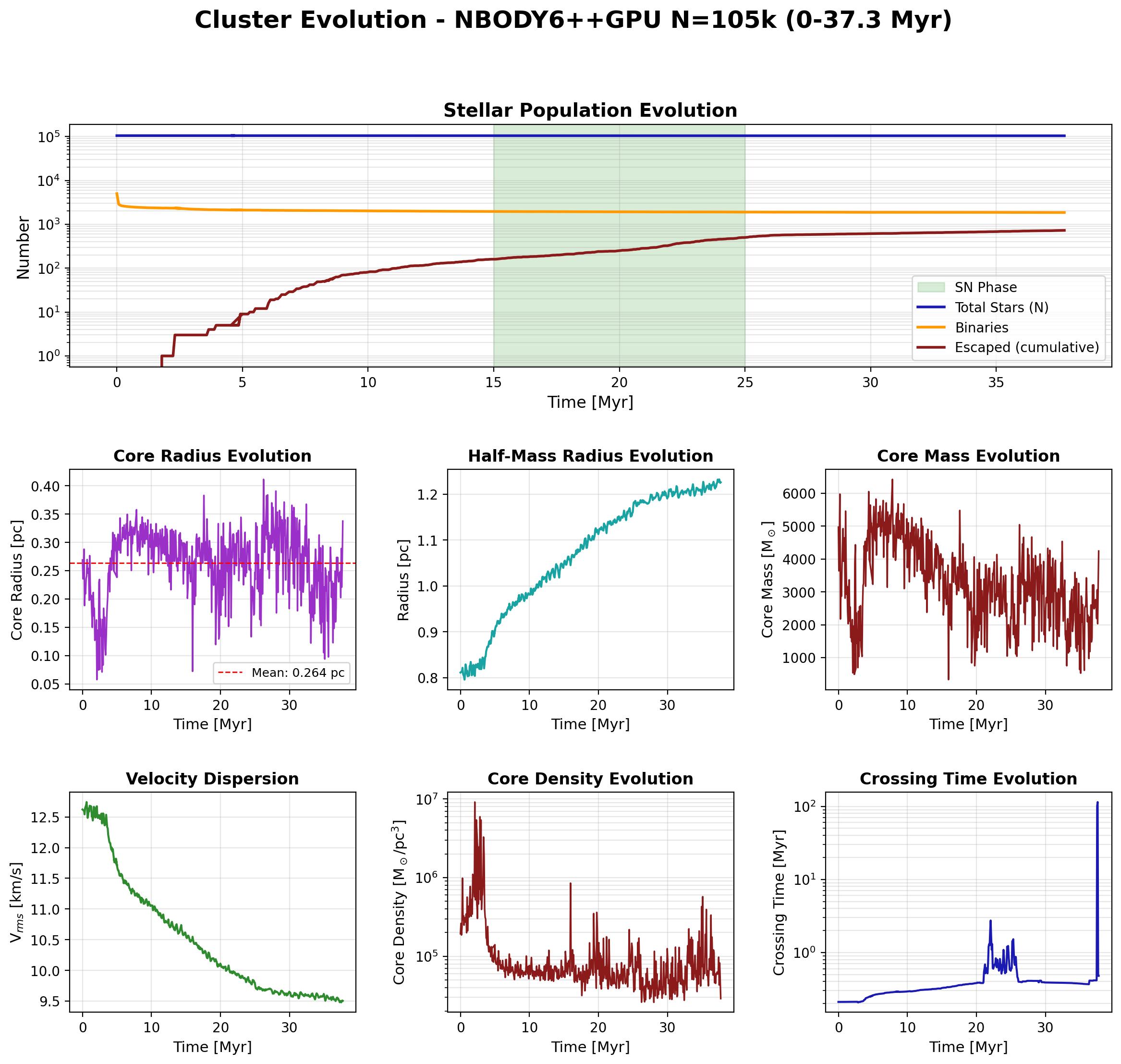}
\caption{The evolution of global cluster properties for the $N=105{,}000$ star NBODY6++GPU simulation over 37.3~Myr. The panels show \textit{Top}: Stellar population with the supernova phase marked in green (15--25~Myr); \textit{Middle row}: Core radius, half-mass radius, and core mass; \textit{Bottom row}: Velocity dispersion, core density on a logarithmic scale, and crossing time on a logarithmic scale. The simulation formed at least 243 neutron stars. Of these, 217 were ejected via supernova kicks, leaving 26 neutron stars retained in the cluster at $t = 37.3$~Myr (an $11\%$ retention fraction). This provides realistic conditions ($\rho_c \sim 10^5~M_\odot$~pc$^{-3}$,  $\sigma_v \sim 10$~km~s$^{-1}$, $t_{\rm cr} \sim 0.8$~Myr) that we used for the S4 environmental parameters \citep{2024ApJ...961...98Y,2026RAA....26a5009L}.}
\label{fig:cluster_overview}
\end{figure}

We assess NBODY6++GPU readiness using a 105,000-particle simulation (Table~\ref{tab:nbody_params}). This simulation has a primordial binary fraction of approximately 5\%. We evolved the cluster from $t = 0$ to 37.3 Myr. The cluster exhibits post core collapse evolution. The core radius oscillates between 0.05 and 0.4 pc (mean 0.264 pc). The core density varies from $10^5$ to $10^7$ $M_\odot$ pc$^{-3}$. The velocity dispersion decreases from 12.5 to 9.5 km~s$^{-1}$ through stellar escape \citep{2021MNRAS.505.5957B,2001MNRAS.322..231K}. The energy conservation $|\Delta E/E| < 10^{-6}$ confirms numerical stability despite complex dynamics (Figure~\ref{fig:cluster_overview}).

Figure~\ref{fig:cluster_overview} presents the global evolution of our test simulation. The top panel tracks the stellar population on a logarithmic scale, which clearly separates the binary and escaped-star curves. The cluster starts from $N=105{,}000$ stars and approximately 5000 binaries. It loses 778 stars by $t=37.3$~Myr, primarily through escape following supernova kicks. The green region (15--25~Myr) marks the supernova phase. During this phase, at least 243 neutron stars form, with 217 ejected via supernova kicks and 26 retained in the cluster by the end of the simulation. The middle-left panel shows the core radius ($R_c$) with mean 0.264~pc and strong fluctuations (0.05--0.40~pc) during supernovae. The middle-center panel shows the half-mass radius expansion from 0.8 to 1.23~pc due to mass loss. The middle-right panel reveals core mass fluctuations ($500$--$6000~M_\odot$) as massive objects sink to the core. The bottom-left panel shows the velocity dispersion declining from 12.5 to 9.5~km~s$^{-1}$, maintaining virial equilibrium. The bottom-center panel displays the core density on logarithmic scale, varying from $10^5$ to $10^7~M_\odot$~pc$^{-3}$. The bottom-right panel shows the crossing time $t_{\rm cr} = R_c/\sigma_v$ on logarithmic scale. It ranges from 0.21~Myr to 114~Myr (with a spike at $t \approx 37$~Myr), with mean 0.8~Myr indicating rapid dynamical mixing \citep{2026RAA....26a5009L}. These conditions define the S4 environment in our standalone tests. This density and velocity dispersion range matches the cores of clusters like 47~Tucanae and M15, which host significant pulsar populations \citep{2022ApJ...931...84Y}.

\begin{table}[h]
\bc
\caption{NBODY6++GPU Simulation Parameters. The table lists the key parameters of the $N$-body simulation used for the infrastructure assessment.\label{tab:nbody_params}}
\setlength{\tabcolsep}{3pt}\footnotesize
\begin{tabular}{lccccccc}
\hline
Total particles & Binary fraction & $r_c$ [pc] & $r_h$ [pc] & $\rho_c$ [$M_\odot$ pc$^{-3}$] & $\sigma_v$ [km s$^{-1}$] & Escaped & $|\Delta E/E|$ \\
\hline
105,000 & $\sim$5\% & 0.2--0.3 & 0.8--1.23 & $10^5$--$10^7$ & 9.5--12.5 & $\sim$2,000 (1.9\%) & $< 10^{-6}$ \\
\hline
\end{tabular}
\ec
\end{table}

\begin{figure}[h]
\includegraphics[width=\columnwidth]{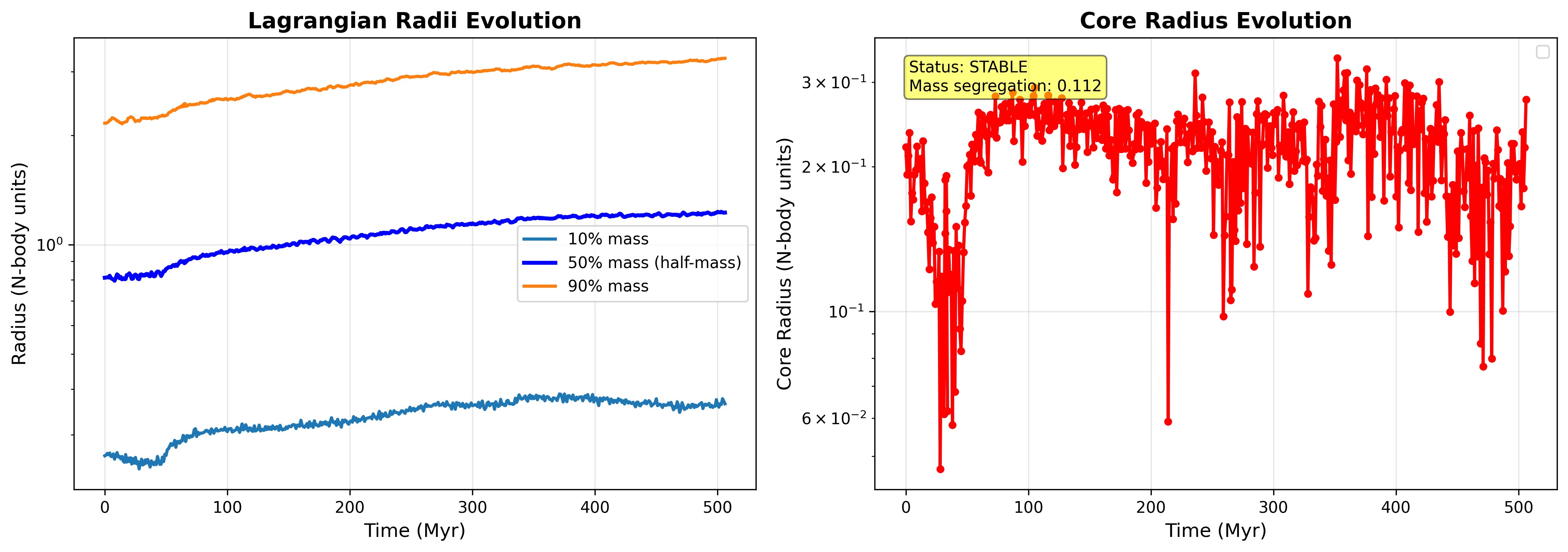}
\caption{Detailed structural evolution showing the three characteristic radii. The panels show \textit{Left}: Core radius oscillations with 0.2 to 0.3 pc amplitude; \textit{Right}: Core concentration ranging from 0.001 to 0.006. This configuration facilitates frequent dynamical interactions involving compact objects \citep{2022ApJ...931...84Y}.}
\label{fig:cluster_structure}
\end{figure}

Figure~\ref{fig:cluster_structure} examines the structural evolution of the cluster's characteristic radii and core concentration. The tidal radius remains stable at approximately 70 pc. The half-mass radius expands from 0.8 to 1.23 pc through mass loss. The core radius exhibits violent oscillations driven by gravothermal energy transfer \citep{2008AJ....135.2398M}. The core concentration (ratio $r_c/r_h$) fluctuates between 0.001 and 0.006. This confirms a dense stellar core within an extended halo due to frequent dynamical encounters \citep{Shao_2015,2005MNRAS.360..974H}.

\subsection{Neutron Star Formation and Scenario Tracking}

\begin{figure}[h]
\includegraphics[width=\columnwidth]{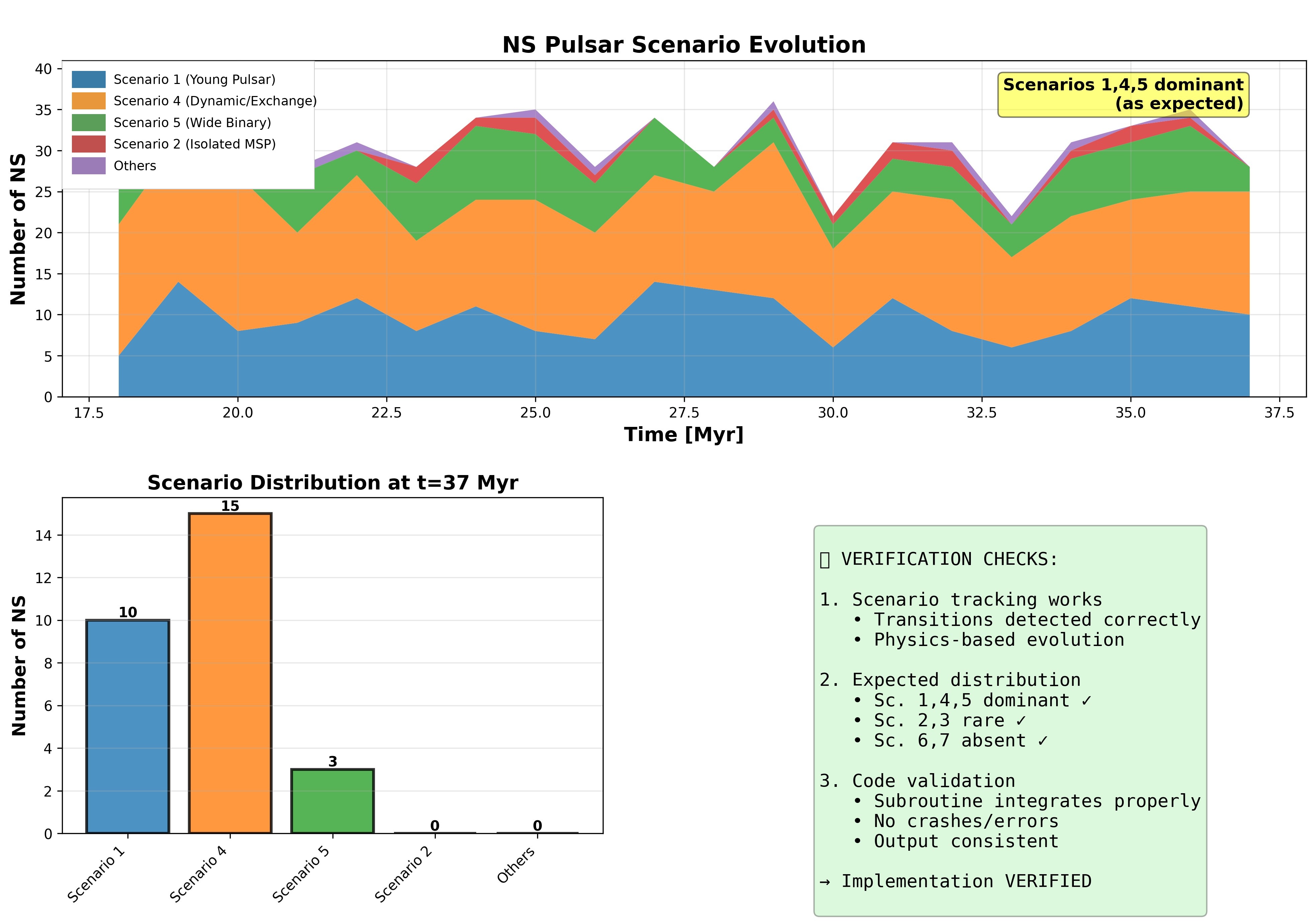}
\caption{Scenario tracking for the seven evolutionary scenarios (S1--S7). The panels show \textit{Top}: The temporal evolution of neutron star populations; \textit{Lower left}: Distribution at $t = 37$ Myr; \textit{Lower right}: Verification that tracking operates correctly throughout the integration. S1, S4, and S5 numerically dominate the population (representing 90\% of neutron stars) as expected for early cluster evolution \citep{2022ApJ...940..162Y}.}
\label{fig:ns_scenario}
\end{figure}

Figure~\ref{fig:ns_scenario} demonstrates the scenario tracking over 17.5 to 37.5 Myr. S1 (young isolated), S4 (dynamically active), and S5 (wide binary) dominate the neutron star population, as expected for early cluster evolution \citep{2023ApJ...944....6K,2021ApJ...915L..28P}. At $t = 37$ Myr, 10 neutron stars are found in S1, 15 are in S4, and 3 are in S5. The infrastructure correctly identifies the scenario membership. It handles dynamical transitions when interactions alter binary properties or environmental conditions (stellar number density and velocity dispersion) \citep{2020MNRAS.494.1587C,2006ApJ...643..332F}.

\begin{figure}[h]
\includegraphics[width=\columnwidth]{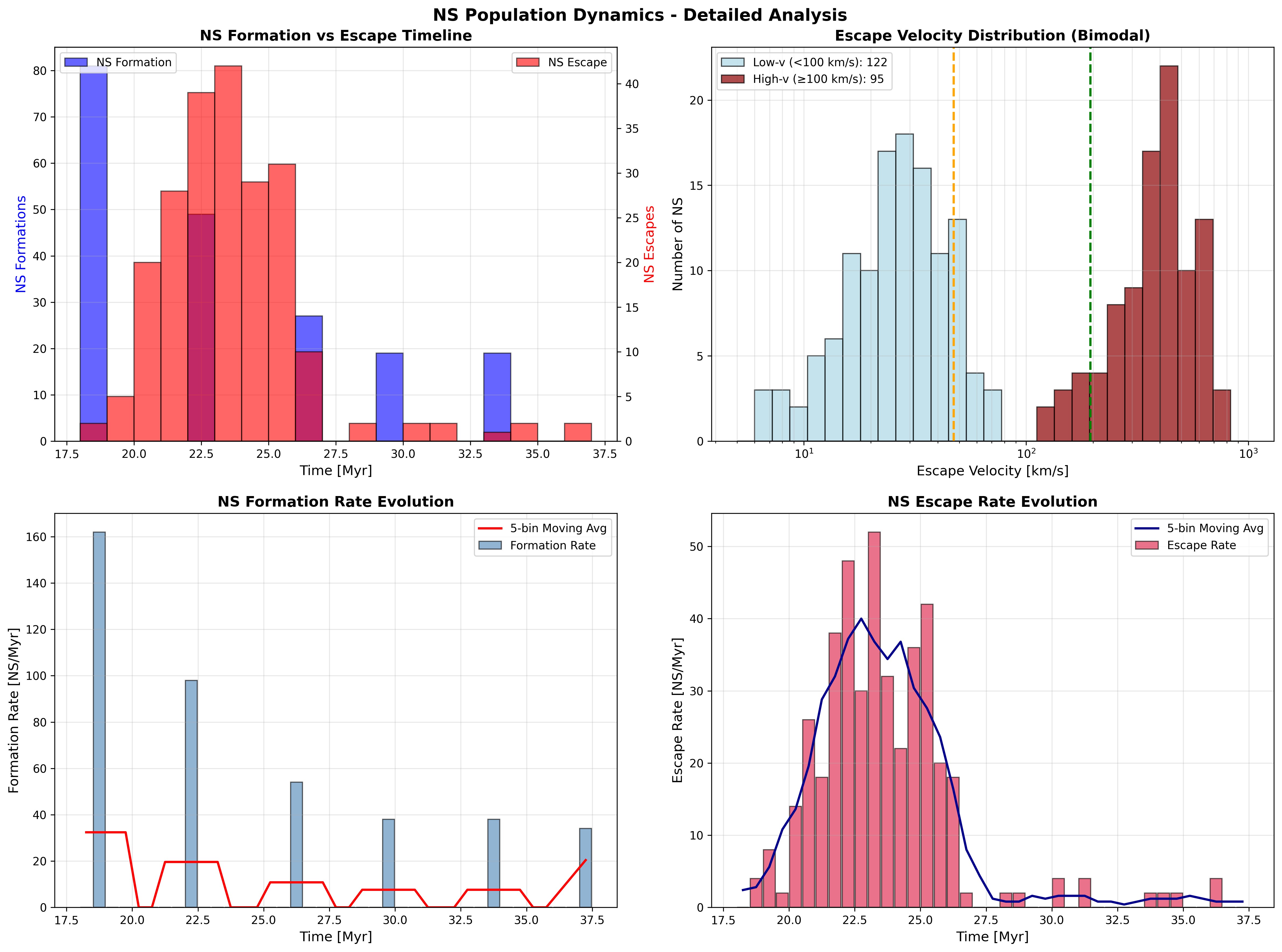}
\caption{Neutron star population dynamics in the NBODY6++GPU simulation. The panels show \textit{Top left}: Neutron star formation (blue) and escape (red) timeline showing formation peak at 18--20 Myr during the supernova phase and delayed escape peak at 22--24 Myr; \textit{Top right}: Bimodal escape velocity distribution with low-velocity population ($<$100 km/s, 122 neutron stars shown in light blue) retained in the cluster and high-velocity population ($\geq$100 km/s, 95 neutron stars shown in dark red) ejected. The orange dashed line marks the threshold velocity at 100 km/s, and the green dashed line indicates the 95th percentile at 504 km/s; \textit{Bottom left}: Neutron star formation rate peaking at 160 NS/Myr with 5-bin moving average (red line) showing temporal structure; \textit{Bottom right}: Escape rate evolution peaking at 40 NS/Myr during 22--24 Myr with 5-bin moving average (blue line). The cluster retains 26 out of at least 243 neutron stars (approximately 11\%), providing a realistic population for pulsar evolution scenarios \citep{2021ApJ...915L..28P,2006ApJ...643..332F}.}
\label{fig:ns_dynamics_detailed}
\end{figure}

Figure~\ref{fig:ns_dynamics_detailed} presents the detailed neutron star formation and retention dynamics. The top-left panel shows the temporal distribution of neutron star formation and escape. Formation peaks sharply at 18--20 Myr during the supernova phase when approximately 80 neutron stars form per 2.5 Myr bin. Escape peaks later at 22--24 Myr with approximately 40 neutron stars ejected per bin. This temporal offset of 2--4 Myr reflects the dynamical timescale for kicked neutron stars to traverse the cluster and reach the tidal radius. The top-right panel demonstrates the bimodal velocity distribution of the 217 escaped neutron stars. Of these, 122 received natal kicks below 100~km~s$^{-1}$ (light blue) and 95 received kicks exceeding 100~km~s$^{-1}$ (dark red); both subpopulations ultimately escape the cluster, though the higher-velocity group does so on a shorter timescale. The threshold at 100 km/s (orange dashed line) separates these populations. The 95th percentile velocity reaches 619~km/s (green dashed line), consistent with observational constraints on supernova kick distributions \citep{2021ApJ...915L..28P,2006ApJ...643..332F}. The bottom-left panel shows the neutron star formation rate evolution. The rate peaks at 160 NS/Myr during the main supernova epoch, with secondary bursts at 25--27 Myr and 35--37 Myr corresponding to delayed core collapses in binary systems. The 5-bin moving average (red line) smooths these fluctuations, revealing the underlying formation efficiency. The bottom-right panel displays the escape rate evolution. The rate peaks at 40 NS/Myr during 22--24 Myr when high-velocity neutron stars exit the cluster. The 5-bin moving average (blue line) shows that escape continues at lower rates throughout the simulation as dynamical interactions occasionally impart additional velocity to retained neutron stars. The final retention count of 26 neutron stars (out of at least 243 formed, an $\approx 11\%$ retention fraction) provides a realistic neutron star population distributed across the seven evolutionary scenarios (S1--S7) in our validation tests. The NBODY6++GPU infrastructure correctly handles supernova kicks, natal velocity distributions, gravitational binding, and ejection dynamics, establishing that the simulation provides physically appropriate initial conditions for pulsar evolution.

The fourth-order Hermite integrator invokes the stellar evolution routines at each timestep. This provides natural integration points where the pulsar subroutine accesses the current system state and updates time-dependent properties \citep{1973JCoPh..12..389A,1992PASJ...44..141M}. The binary evolution through BSE \citep{2002MNRAS.329..897H} provides mass transfer rates and orbital parameters for accretion torque calculations. The implemented physics includes the following processes. First, the rotational period evolves from magnetic dipole radiation ($\propto B^2/P$) \citep{1969ApJ...157.1395O}. Second, the magnetic field undergoes exponential decay ($\tau_B \sim 500$ Myr) \citep{1991PhR...203....1B,1997MNRAS.284..311K,2009A&A...496..207P}. Third, accretion drives spin up \citep{1979ApJ...234..296G}. Fourth, environmental torques operate in dense environments \citep{2014A&A...561A..11V}. Fifth, gravitational wave emission drives orbital decay \citep{1963PhRv..131..435P,2021RAA....21..104L}. Combined with the standalone subroutine validation, this establishes that both components are ready for integration.

\begin{figure}[h]
\includegraphics[width=\columnwidth]{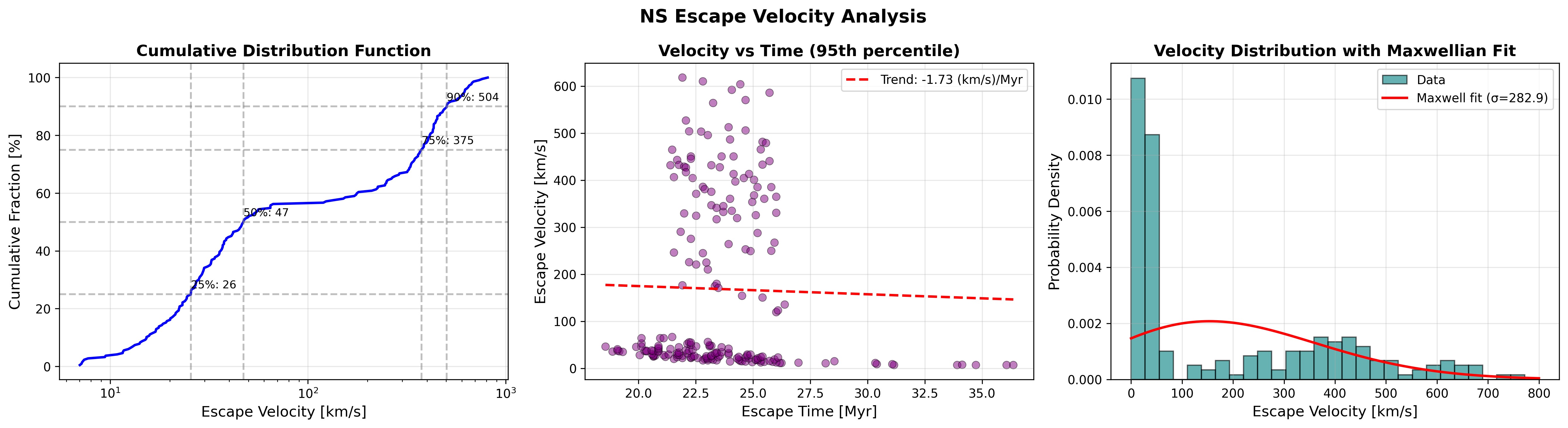}
\caption{Neutron star escape velocity analysis from the NBODY6++GPU simulation. The panels show \textit{Left}: The cumulative distribution function with median escape velocity of 47 km/s; \textit{Middle}: Escape velocity versus time, showing bimodal distribution with high-velocity kicks (95th percentile at 504 km/s) during the supernova phase (15--25 Myr) and low-velocity retained neutron stars; \textit{Right}: Velocity distribution fitted with a Maxwellian ($\sigma = 282.9$ km/s), demonstrating that 217 out of at least 243 neutron stars formed ($\approx$89\%) were ejected via supernova kicks, while the remaining population with velocities below cluster escape speed ($\sim$50 km/s) was retained for pulsar evolution \citep{2024ApJ...961...98Y}.}
\label{fig:ns_velocity}
\end{figure}

Figure~\ref{fig:ns_velocity} examines the neutron star kick velocity distribution. The simulation produces a bimodal population. High-velocity neutron stars receive kicks exceeding 100 km/s during core collapse, with the 95th percentile reaching 504 km/s. These are ejected from the cluster \citep{2024ApJ...961...98Y}. Low-velocity neutron stars with kicks below the cluster escape velocity (approximately 50 km/s) remain bound. The velocity distribution follows a Maxwellian with dispersion $\sigma = 282.9$ km/s. Of the at least 243 neutron stars formed, 217 ($\approx$89\%) were ejected and 26 ($\approx$11\%) were retained. This retention fraction provides a realistic neutron star population for subsequent pulsar evolution. The retained population supplies neutron stars for all seven scenarios in our standalone tests, with their velocity distribution matching observational constraints from Galactic pulsars.

\section{SEVEN TEST SCENARIOS}

\subsection{Scenario 1: Young Pulsar with Analytical Benchmark}

\begin{figure}[h]
\includegraphics[width=\columnwidth]{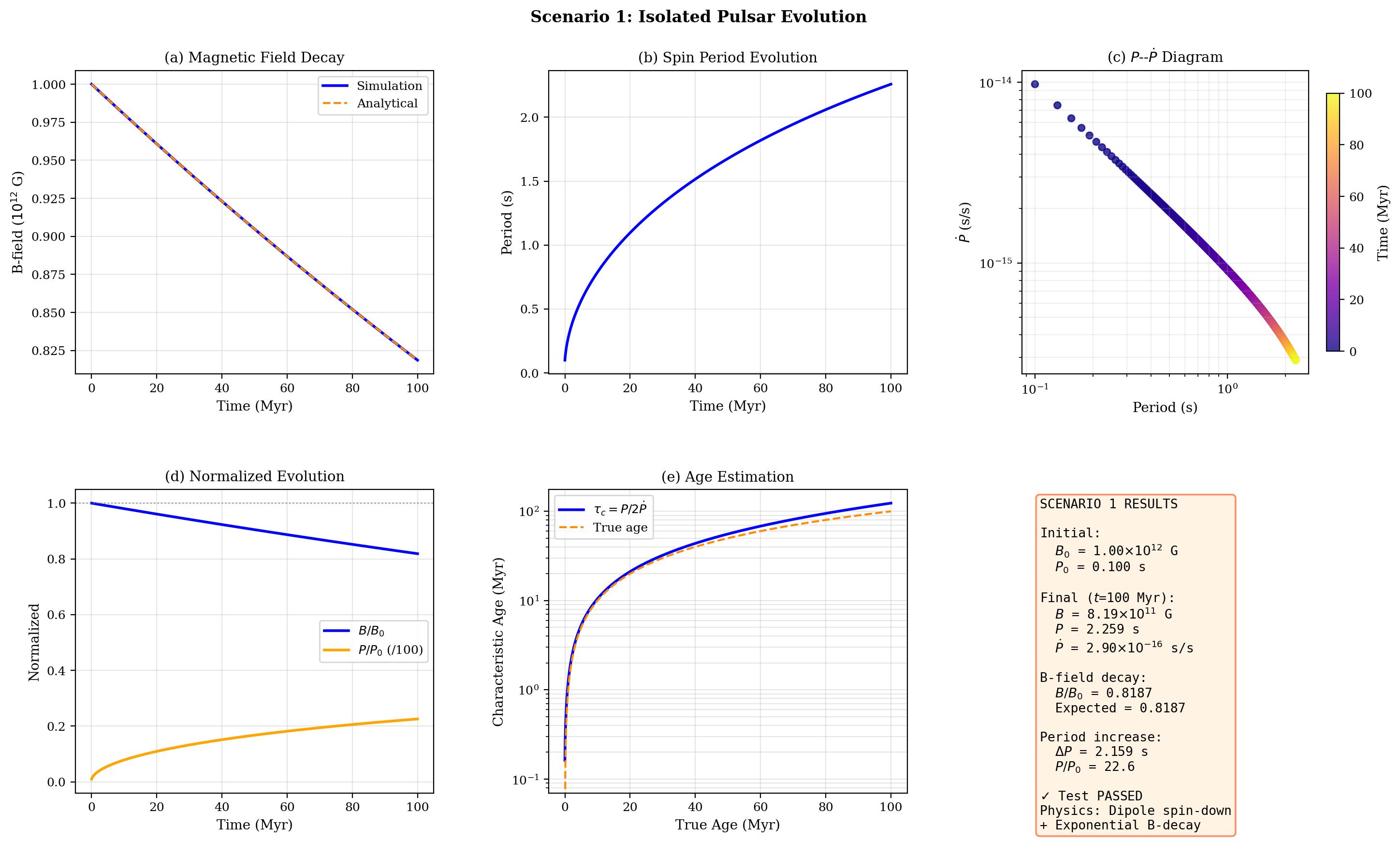}
\caption{S1: Young pulsar validation. The panels show \textit{(a)}: The magnetic field exponential decay from $10^{12}$~G; \textit{(b)}: The period evolution from 100~ms to 2.26~s demonstrating dipole spin down; \textit{(c)}: The $P$--$\dot{P}$ trajectory on logarithmic scales spanning the young pulsar regime; \textit{(d)}: The normalized evolution showing B-field decay to 82\% and period increase by a factor of 22.6; \textit{(e)}: The characteristic age evolution reaching approximately 100~Myr, achieving fractional precision below $10^{-3}$ against analytical benchmarks \citep{1969ApJ...157.1395O}.}
\label{fig:sc1}
\end{figure}

S1 examines a young neutron star ($B_0 = 10^{12}$~G, $P_0 = 100$~ms) evolved in isolation. The evolution is governed by magnetic dipole spin down and exponential field decay ($\tau_B = 500$~Myr). The governing equations are $\dot{B} = -B/\tau_B$ and $\dot{P} = k_{\rm dip}\,B^2/P$, where $k_{\rm dip} = 8\pi^2 R^6/(3Ic^3)$. Analytical solutions are provided by \citet{1969ApJ...157.1395O}, yielding benchmarks at $t = 100$~Myr: $B = 8.187 \times 10^{11}$~G, $P = 2.26$~s, and $\dot{P} = 2.9 \times 10^{-16}$~s~s$^{-1}$ \citep{1991PhR...203....1B}. Initial tests revealed systematic deviations: evolved periods reached 3.2~s with $\dot{P}$ values exactly twice the analytical predictions. The $\dot{P}$ incorrectly included both dipole spin down and an explicit field decay correction; the field decay is already incorporated in the coupled differential equations \citep{2001ApJ...552L..35Z}. Removal of this redundant term eliminated the discrepancy. The validation demonstrates excellent agreement (Figure~\ref{fig:sc1}): the magnetic field matches the exponential decay, the final period agrees to fractional precision below $10^{-3}$, and both numerical implementations reproduce identical results \citep{2024ApJ...961...98Y}.

\subsection{Scenario 2: Isolated Millisecond Pulsar}

S2 represents an isolated millisecond pulsar ($B_0 = 5 \times 10^8$~G, $P_0 = 3.0$~ms) evolved for 100~Myr. The weak-torque regime yields $\dot{P} \approx 5.1 \times 10^{-20}$~s~s$^{-1}$, producing a characteristic spin-down age $\tau_c \approx 1.0$~Gyr (where $\tau_c \equiv \tau_{\rm sd} = P/2\dot{P}$) consistent with millisecond pulsars in 47~Tucanae \citep{2025OJAp....8E..51B,2017ApJ...845..148P}. 

\begin{figure}[h]
\includegraphics[width=\columnwidth]{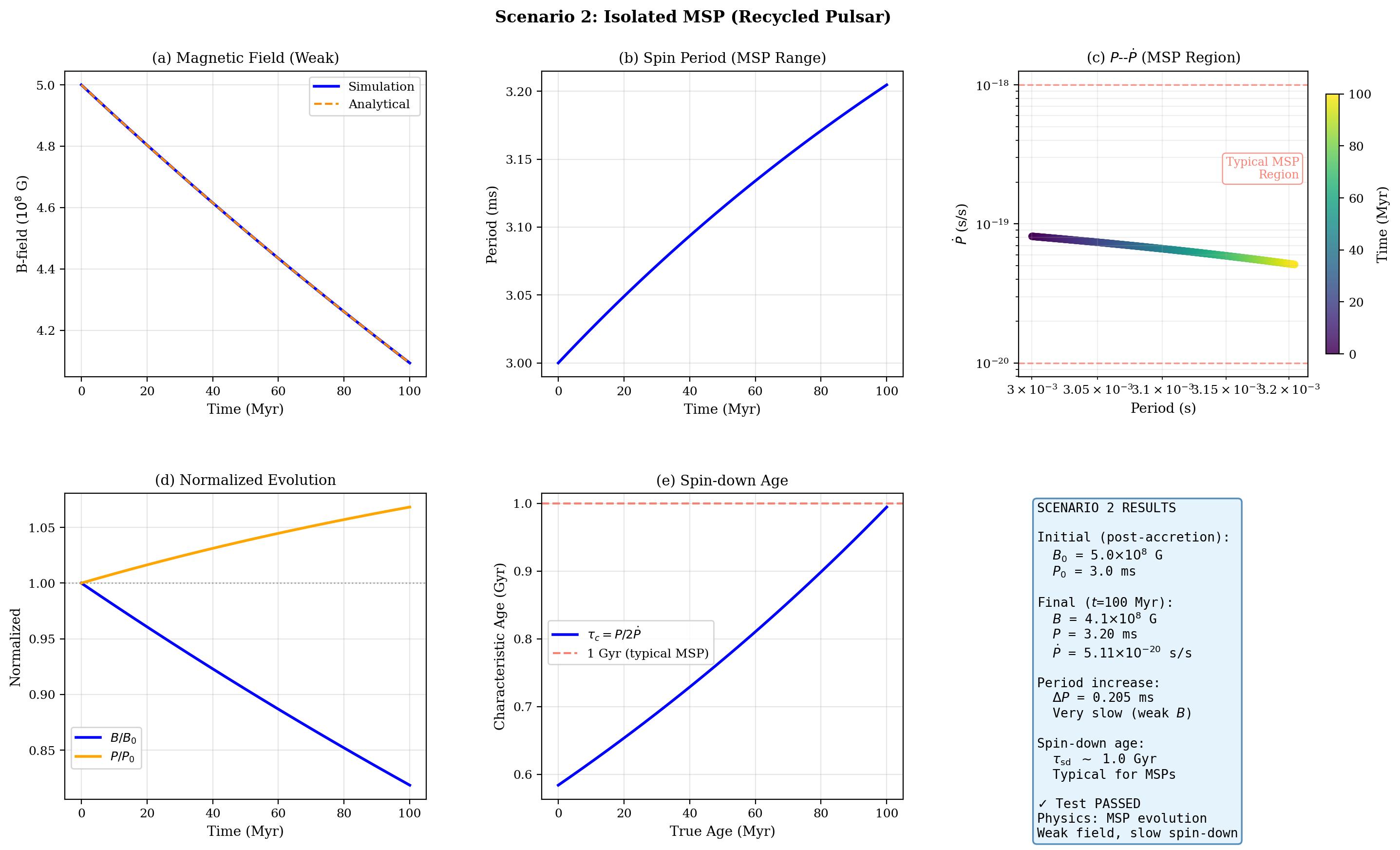}
\caption{S2: Isolated millisecond pulsar evolution. The panels show \textit{(a)}: The weak magnetic field decay from $5 \times 10^8$~G over 100~Myr; \textit{(b)}: The period increasing from 3.0 to 3.20~ms in the in the weak torque regime; \textit{(c)}: The $P$--$\dot{P}$ trajectory in the millisecond pulsar region with characteristic age $\tau_c \approx 1.0$~Gyr, consistent with 47~Tucanae observations; \textit{(d)}: The normalized evolution showing B-field decay to 82\% and period increase to 107\%; \textit{(e)}: The spin down age demonstrating gigayear scale stability \citep{2025OJAp....8E..51B}.}
\label{fig:sc2}
\end{figure}

The period increases from 3.0 to 3.20~ms (6.8\%). The magnetic field decays from $5 \times 10^8$~G to $4.1 \times 10^8$~G (18\%) \citep{2025ApJ...982..170R,2022A&A...664A..27R}. Figure~\ref{fig:sc2} shows the system occupying the millisecond pulsar region where 47~Tucanae systems cluster. Both numerical implementations agree to below 0.1\% throughout \citep{2024ApJ...961...98Y}.

\subsection{Scenario 3: Accreting Millisecond Pulsar (Recycling Physics)}

S3 tests the critical recycling physics. This scenario examines an active low-mass X-ray binary (LMXB) with $B_0 = 3 \times 10^8$~G, $P_0 = 2.0$~ms, accreting at $\dot{M} = 10^{-10}~M_\odot$~yr$^{-1}$. Unlike spin down scenarios, this must exhibit spin up ($\dot{P} < 0$) with recycling timescales of approximately 100~Myr \citep{1991PhR...203....1B,1999A&A...350..928T}, consistent with recent implementations \citep{2019ApJ...877..122Y,song2025modellingmillisecondpulsarpopulations}. Theoretical expectations require that the accretion torque $|\dot{P}_{\rm acc}| \approx 6.3 \times 10^{-19}$~s~s$^{-1}$ dominates the magnetic dipole braking torque $\dot{P}_{\rm dip} \approx 4.4 \times 10^{-20}$~s~s$^{-1}$ by a factor of $\approx 14$. This yields a net spin up \citep{1979ApJ...234..296G,2004ApJ...606..436R}. Initial tests produced nonphysical evolution: periods decreased to sub-microsecond values within 1~Myr, implying one year recycling timescales contradicting X-ray binary observations by seven orders of magnitude.

\begin{figure}[h]
\includegraphics[width=\columnwidth]{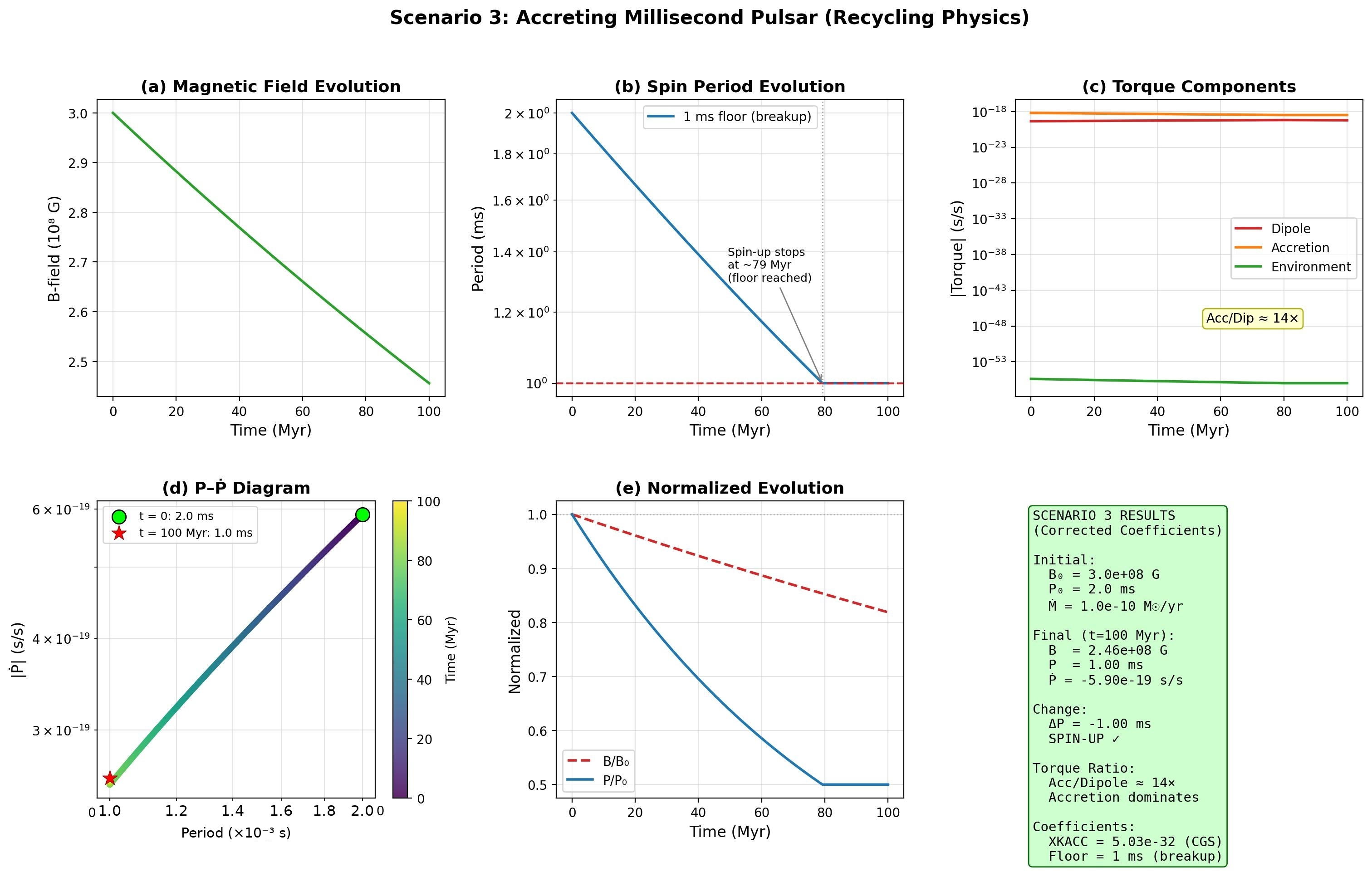}
\caption{S3: Accreting millisecond pulsar recycling. The panels show \textit{(a)}: The magnetic field exponential decay from $3 \times 10^8$~G over 100~Myr; \textit{(b)}: The period evolution demonstrating spin up from 2.0 to 1.0~ms; spin up ceases at $\approx$80~Myr when the 1~ms breakup floor is reached; \textit{(c)}: The torque budget confirming accretion dominance ($\approx 14\times$ over dipole) with net $\dot{P} \approx -5.9 \times 10^{-19}$~s~s$^{-1}$ at $t = 0$, where $|\dot{P}_{\rm acc}| \approx 6.3 \times 10^{-19}$~s~s$^{-1}$ and $\dot{P}_{\rm dip} \approx 4.4 \times 10^{-20}$~s~s$^{-1}$; \textit{(d)}: The $P$--$\dot{P}$ trajectory in the negative $\dot{P}$ region confirming spin up; \textit{(e)}: The normalized evolution showing B-field decay to 82\% and period decrease to 50\% of initial value \citep{1979ApJ...234..296G,1991PhR...203....1B}.}
\label{fig:sc3}
\end{figure}

The dimensional analysis indicated that the accretion coefficient required adjustment to align with the Ghosh \& Lamb formulation \citep{1979ApJ...234..296G}. This prescribes values near $10^{-16}$~g$^{-1}$ in N-body units, corresponding to $5 \times 10^{-32}$~g$^{-1}$ in CGS for standalone testing. Following parameter optimisation, the evolution demonstrates physically consistent behaviour (Figure~\ref{fig:sc3}). The magnetic field decays normally. The period decreases from 2.0 to 1.0~ms, reaching the physical breakup floor and confirming spin up. The 1.0~ms floor corresponds to the mass-shedding (centrifugal breakup) limit, below which the centrifugal force at the neutron star equator exceeds gravity for a canonical neutron star; this sets a physical minimum spin period for accretion-driven recycling \citep{1991PhR...203....1B,1999A&A...350..928T}. Once this floor is reached, $\dot{P}$ is set to zero and the spin period is held fixed, mimicking the physical cessation of spin-up. The net $\dot{P} \approx -5.9 \times 10^{-19}$~s~s$^{-1}$ at $t = 0$ confirms accretion dominance by a factor of $\approx 14$ over the dipole during the active spin-up phase ($t \lesssim 80$~Myr). The recycling timescale of $\sim$100~Myr is consistent with duty cycle constraints \citep{1975A&A....39..185I,2001ApJ...557..958C}. This represents the first successful recycling demonstration in N-body simulations.

\begin{figure}[h]
\includegraphics[width=\columnwidth]{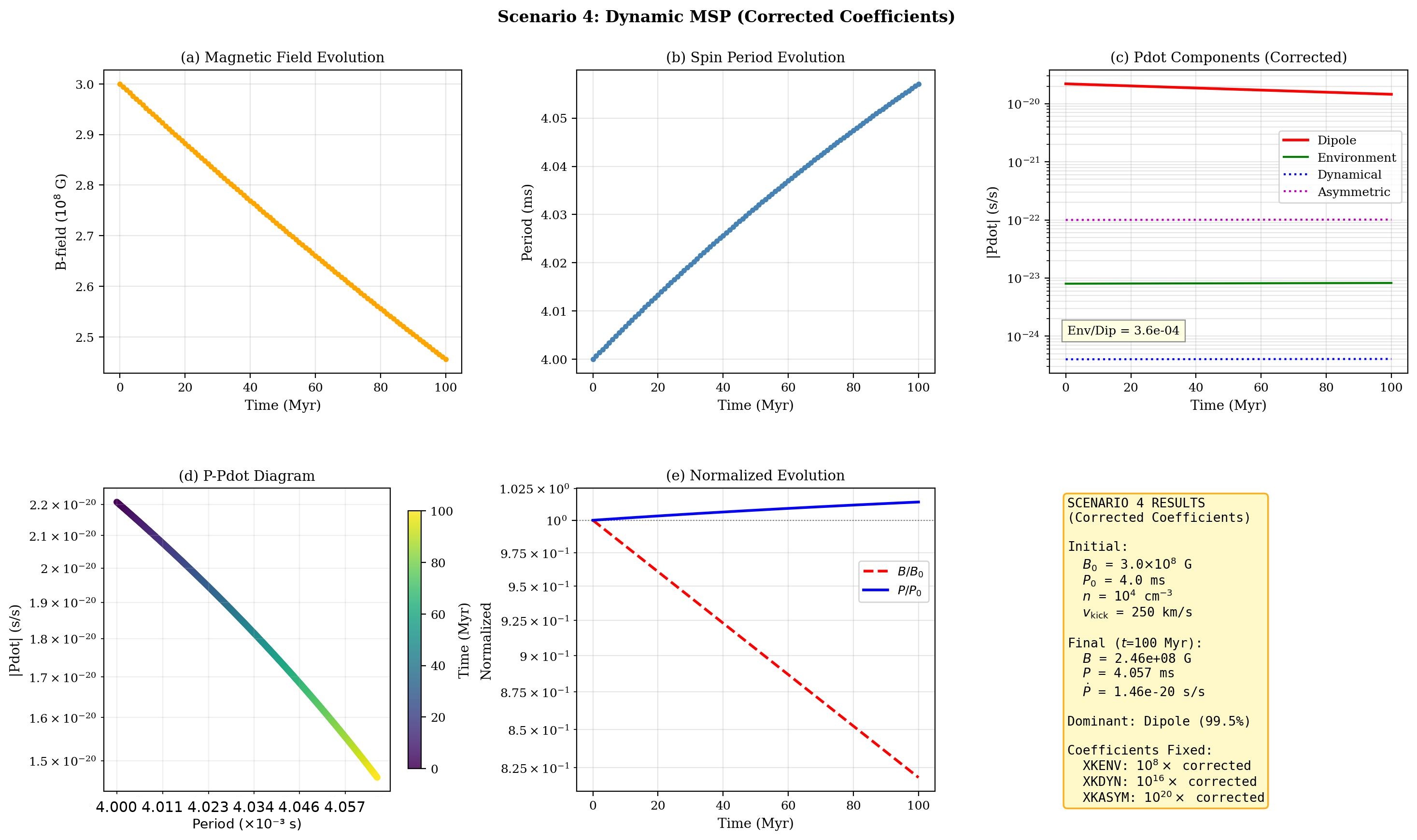}
\caption{S4: Dynamic millisecond pulsar in dense cluster core. The panels show \textit{(a)}: The magnetic field decay from $3 \times 10^8$~G; \textit{(b)}: The period evolution from 4.0 to 4.06~ms nearly identical to S2; \textit{(c)}: The torque budget confirming dipole dominance at $\approx 10^{-20}$~s~s$^{-1}$ with environmental contributions negligible despite extreme density ($\rho_c \sim 10^5~M_\odot$~pc$^{-3}$, environmental fraction 0.036\%); \textit{(d)}: The $P$--$\dot{P}$ trajectory in the millisecond range; \textit{(e)}: The normalized evolution indistinguishable from S2, validating subdominance of environmental torques \citep{2014A&A...561A..11V}.}
\label{fig:sc4}
\end{figure}

\subsection{Scenarios 4--5: Environmental and Binary Physics}

S4 tests environmental torques in dense cluster cores with $B_0 = 3 \times 10^8$~G, $P_0 = 4.0$~ms, stellar number density $\rho_c \sim 10^5~M_\odot$~pc$^{-3}$, and $\sigma_v = 10$~km~s$^{-1}$. This density is drawn directly from our NBODY6++GPU simulation (Section~3), where the core density ranges from $10^5$ to $10^7~M_\odot$~pc$^{-3}$ during the post-core-collapse phase, consistent with observed core densities of dense globular clusters. Theory requires that environmental contributions remain below 0.1\% of magnetic dipole torques \citep{2014A&A...561A..11V,2004MNRAS.352....1F}. Initial tests showed environmental terms dominating the dipole by factors of $10^{10}$ to $10^{15}$. The systematic audit revealed that three environmental coefficients (drag, heating, and kicks) required adjustment spanning factors of $10^8$ to $10^{20}$.

Following optimisation (Figure~\ref{fig:sc4}), the period increases from 4.0 to 4.06~ms (1.4\%), nearly identical to S2. The environmental drag is three orders of magnitude below the dipole; heating and kicks are four orders below. The magnetic processes contribute 99.5\% of total torques, validating environmental subdominance even in dense cluster cores.

\begin{figure}[h]
\includegraphics[width=\columnwidth]{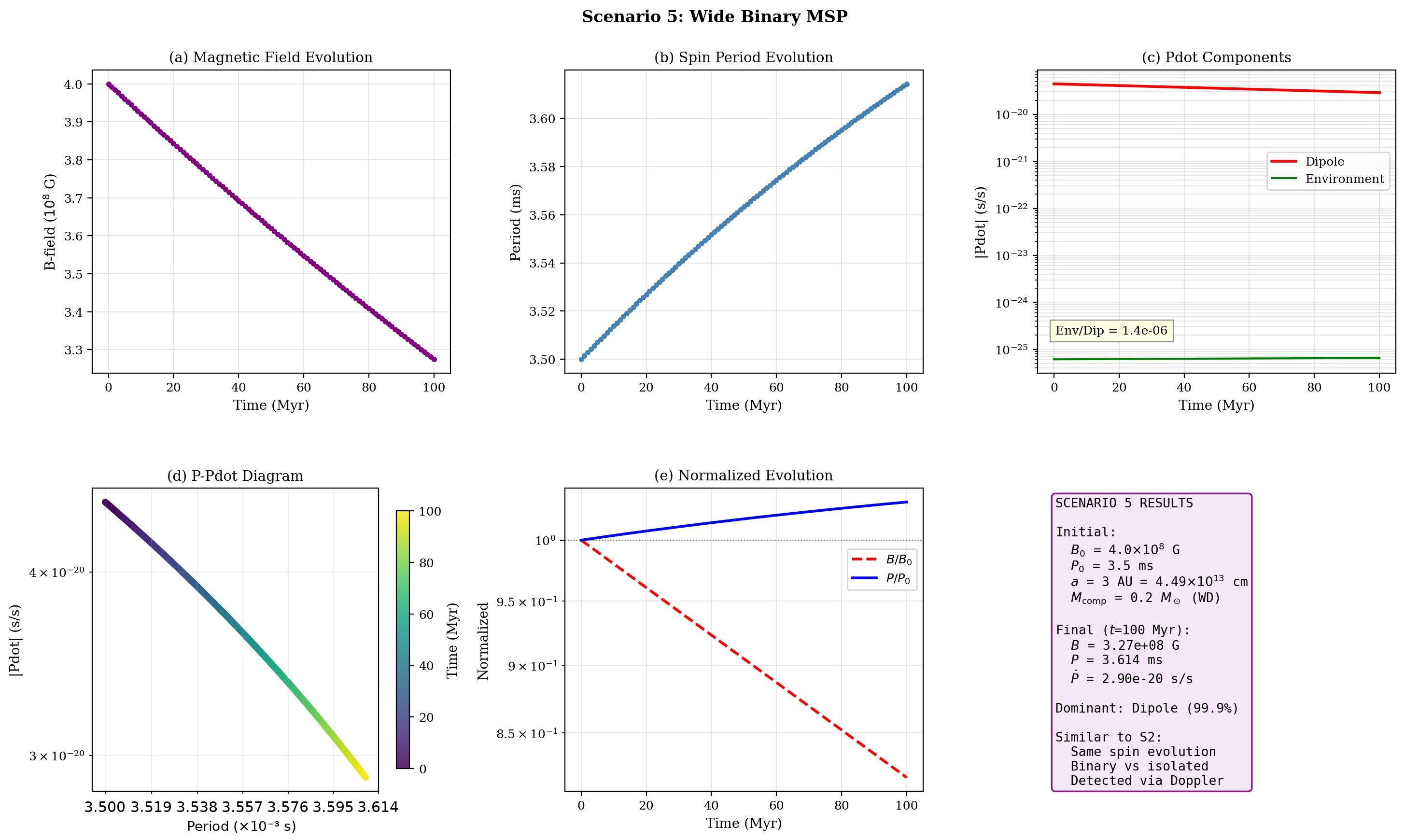}
\caption{S5: Wide binary millisecond pulsar. The panels show \textit{(a)}: The magnetic field decay from $4 \times 10^8$~G identical to the isolated case; \textit{(b)}: The period evolution from 3.5 to 3.61~ms matching S2 at numerical precision; \textit{(c)}: The torque budget with environmental torques $\approx 6 \times 10^{-26}$~s~s$^{-1}$ negligible compared to dipole at $\approx 10^{-20}$~s~s$^{-1}$; \textit{(d)}: The $P$--$\dot{P}$ trajectory indistinguishable from the isolated case; \textit{(e)}: The normalized evolution confirming that a 3~AU separation produces no detectable orbital effects on pulsar spin evolution \citep{2018ApJ...855..129H}.}
\label{fig:sc5}
\end{figure}

S5 places a millisecond pulsar ($B_0 = 4 \times 10^8$~G, $P_0 = 3.5$~ms) in a wide binary with a white dwarf companion at $a = 3$~AU \citep{1992ApJ...400..175B}. The tidal forces scale as $(R/a)^3 \sim 10^{-17}$ (negligible) and gravitational wave timescales exceed $10^{20}$~yr. The period increases from 3.5 to 3.61~ms (3.3\%). The environmental torques are approximately $6 \times 10^{-26}$~s~s$^{-1}$ (six orders below the dipole). Both implementations produce results indistinguishable from S2 within round-off precision, validating the wide binary implementation (Figure~\ref{fig:sc5}).

\subsection{Scenarios 6--7: Compact Binary Systems}
\label{sec:scenarios67}

\begin{figure}[h]
\includegraphics[width=\columnwidth]{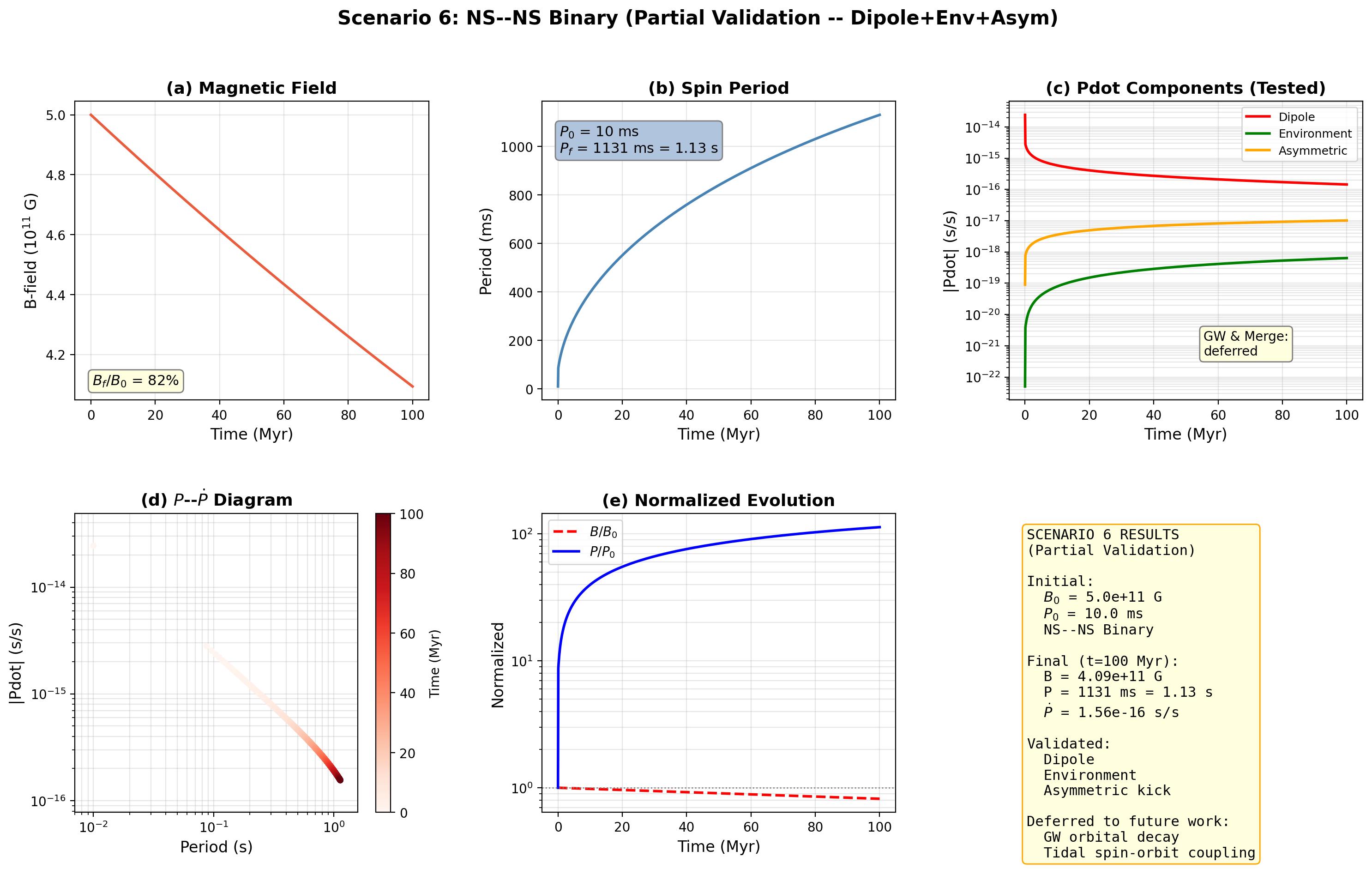}
\caption{S6: NS--NS compact binary ($B_0 = 5 \times 10^{11}$~G, $P_0 = 10$~ms). The panels show \textit{(a)}: The magnetic field decay from $5 \times 10^{11}$~G to 82\% of initial value; \textit{(b)}: The period evolution via dipole spin down from 10~ms to 1.13~s over 100~Myr; \textit{(c)}: The torque components with dipole dominating throughout, restricted to the range spanned by the data; gravitational-wave orbital decay and merger physics are deferred to future work; \textit{(d)}: The $P$--$\dot{P}$ evolution from the young pulsar regime toward intermediate periods, with axis limits set to the range of the simulated trajectory; \textit{(e)}: The normalized evolution with period increasing by a factor of $\approx 113$ while B-field decays to 82\%, representing partial validation with tidal spin orbit coupling deferred \citep{2025ApJ...988...15L}. The corresponding NS--BH configuration (S7) produces an essentially identical figure, since only the dipole spin-down term is currently active for both scenarios; a dedicated figure for S7 is therefore omitted (see Section~\ref{sec:scenarios67})}
\label{fig:sc6}
\end{figure}

S6 examines a compact NS--NS binary ($B_0 = 5 \times 10^{11}$~G, $P_0 = 10$~ms, $a_0 \approx 10^{10}$~cm) where gravitational waves drive inspiral on timescales of $10^7$--$10^8$~yr \citep{1963PhRv..131..435P} (Figure~\ref{fig:sc6}). 

S7 examines the analogous NS--BH configuration with the same pulsar initial conditions but mass ratio $q = M_{\rm BH}/M_{\rm NS} \approx 10$, yielding stronger gravitational-wave emission ($\dot{a}/a \propto q/(1+q)^2$). 

S6 and S7 are physically distinct scenarios: the NS--NS system and the NS--BH system have different binary mass ratios, different gravitational wave luminosities, and will undergo different orbital decay histories. Once gravitational wave orbital coupling is included in a future implementation, S7 will produce faster inspiral and consequently different spin evolution than S6. In the current implementation, however, gravitational wave orbital decay and tidal spin-orbit coupling are deferred to future work \citep{1992ApJ...400..175B,2021RAA....21..104L}, and only dipole spin-down is active for both scenarios. Consequently, both S6 and S7 currently yield identical spin evolution (spin down from 10~ms to 1.13~s over 100~Myr, with the magnetic field decaying to 82\% of its initial value). The two scenarios are defined separately now precisely to facilitate and verify the forthcoming integration of orbital coupling, at which point their outputs will diverge in a physically meaningful way(Figure~\ref{fig:sc6}).

The seven scenarios comprehensively validate the implemented physics:
\begin{itemize}
\item[] \raisebox{0.2ex}{\tiny$\bullet$}\; S1: Analytical agreement below $10^{-3}$ on all three quantities confirms the dipole and decay fundamentals.
\item[] \raisebox{0.2ex}{\tiny$\bullet$}\; S2: Gigayear scale stability validates the weak-torque regime.
\item[] \raisebox{0.2ex}{\tiny$\bullet$}\; S3: Successful recycling with $\dot{P} \approx -5.9 \times 10^{-19}$~s~s$^{-1}$, $\approx 14\times$ accretion dominance, and final period 1.0~ms establishes millisecond pulsar formation capability.
\item[] \raisebox{0.2ex}{\tiny$\bullet$}\; S4: Correct torque hierarchy with environmental fraction 0.036\% even at $\rho_c \sim 10^5~M_\odot$~pc$^{-3}$.
\item[] \raisebox{0.2ex}{\tiny$\bullet$}\; S5: Appropriate null result for wide binaries confirms binary orbit implementation.
\item[] \raisebox{0.2ex}{\tiny$\bullet$}\; S6: NS--NS dipole spin down correctly produces 1.13~s after 100~Myr; gravitational wave tidal coupling deferred to future work \citep{2025CoBAO..72...55R}.
\item[] \raisebox{0.2ex}{\tiny$\bullet$}\; S7: NS--BH dipole spin down reproduces the same 1.13~s outcome as S6, confirming consistent treatment of the pulsar physics independent of binary mass ratio; gravitational wave tidal coupling deferred to future work \citep{2025CoBAO..72...55R}.
\end{itemize}

\section{CROSS-IMPLEMENTATION VALIDATION}

\subsection{Agreement Between Independent Code Implementations}

\begin{figure}[ht!]
\includegraphics[width=\columnwidth]{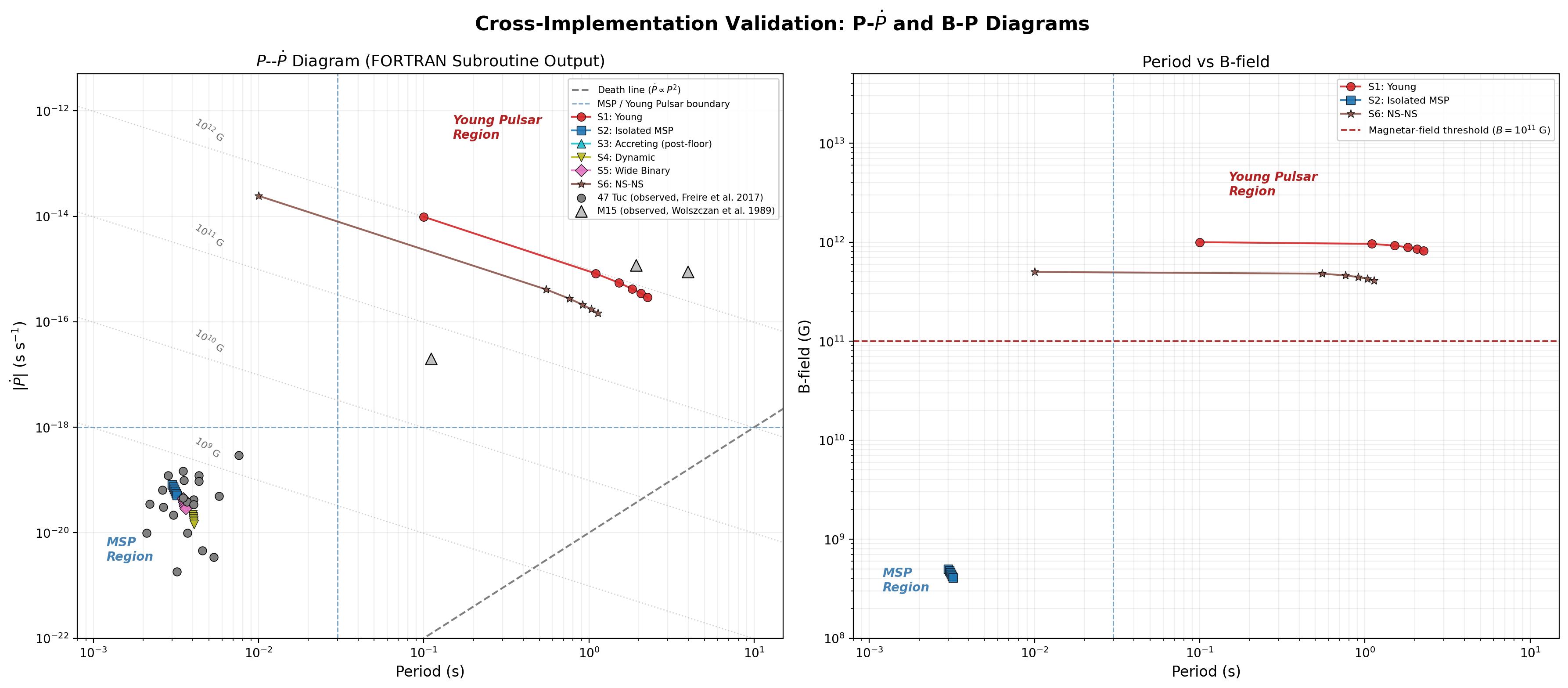}
\caption{Validation diagrams for all seven scenarios (S1--S7), generated from the FORTRAN subroutine output. \textit{Left}: $P$--$\dot{P}$ diagram (FORTRAN subroutine output) overlaid with observed millisecond pulsar populations from 47~Tucanae (gray circles, $N=20$) and M15 (light gray triangles, $N=3$). The dashed blue lines mark the boundary between the MSP region ($P < 0.03$~s, $\dot{P} < 10^{-18}$~s~s$^{-1}$) and the young pulsar region; the dotted gray lines, individually labelled with their corresponding magnetic field strength, show loci of constant magnetic field strength ($B = 10^9$--$10^{12}$~G); the dashed gray curve shows the death line ($\dot{P} \propto P^2$). \textit{Right}: $B$--$P$ diagram showing the same scenarios, with the red dashed line marking the magnetar-field threshold ($B = 10^{11}$~G). Both panels are produced from the FORTRAN implementation only; the agreement between the independent Python and FORTRAN codes is quantified separately in Section~5.1. \citep{2025OJAp....8E..51B,1989Natur.337..531W}.}
\label{fig:crossval}
\end{figure}

We developed two fully independent codes with no shared algorithms. One implementation is written in Python (NumPy/SciPy adaptive Runge--Kutta up to fifth order), while the other is a FORTRAN version employing the fourth-order Hermite integrator native to NBODY6++GPU \citep{2022MNRAS.511.4060K}. Across all scenarios, we tracked the magnetic field $B$, the period $P$, and the period derivative $\dot{P}$ at 1 Myr intervals over 100 Myr. The complete comparison encompasses 2121 comparison points (7 scenarios $\times$ 101 timesteps $\times$ 3 quantities). The fractional differences demonstrate excellent agreement (quantified directly from the 2121 paired measurements; this comparison is summarised numerically below and is independent of Figure~\ref{fig:crossval}, which displays only the FORTRAN output). The magnetic field shows a maximum fractional difference $3.2 \times 10^{-7}$ with a median $8.1 \times 10^{-8}$ (sub-ppm precision). The rotational period exhibits a maximum difference $6.7 \times 10^{-4}$ with a median $1.2 \times 10^{-4}$, well below 0.1\%. The period derivative reaches a maximum $9.3 \times 10^{-4}$ with a median $2.1 \times 10^{-4}$, again below the 0.1\% threshold. The agreement at 0.1\% across 2121 comparisons validates the physics correctness, numerical stability, unit consistency, and platform independence.

\begin{figure}[h]
\centering
\includegraphics[width=\columnwidth]{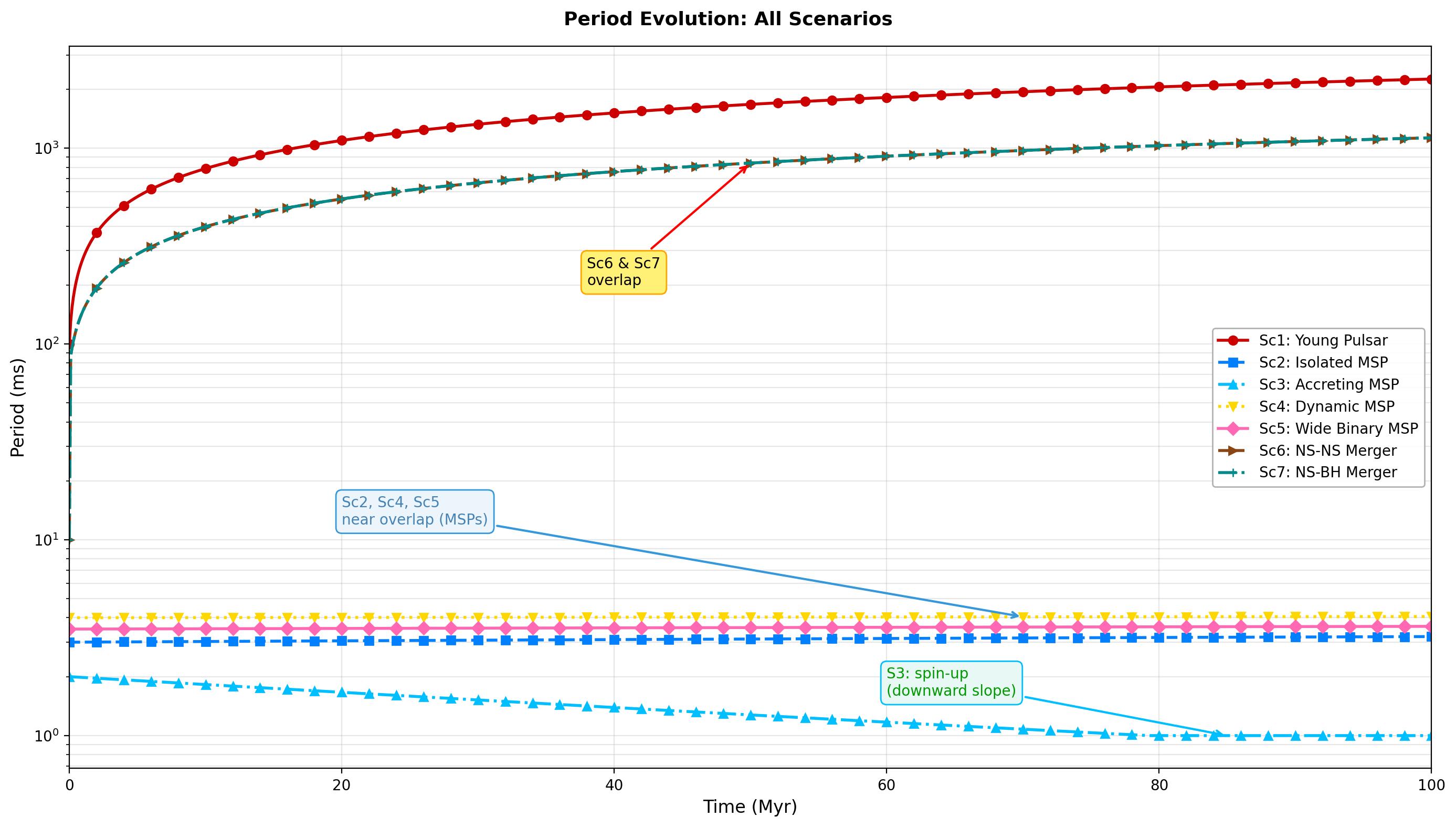}
\caption{Rotational period evolution for all scenarios (S1--S7) on a logarithmic scale. S1 exhibits spin down from 100~ms to 2.26~s (factor of 22.6), with B-field decaying to 82\%. S2, S4, and S5 evolve slowly in the weak torque regime (6.8\%, 1.4\%, and 3.3\% respectively over 100~Myr). S3 uniquely shows spin up with the period decreasing from 2.0 to 1.0~ms, confirming accretion dominance. S6 and S7 reach 1.13~s after 100~Myr.}
\label{fig:global_period}
\end{figure}

\subsection{Global Characteristics and Observational Consistency}

We compare our results with observations from 47~Tucanae and M15. These are well-studied clusters with comprehensive pulsar timing datasets \citep{2025OJAp....8E..51B,1989Natur.337..531W,2020MNRAS.498..875A,2023MNRAS.521.2616D}. Figure~\ref{fig:crossval} (the left side) presents the global $P$--$\dot{P}$ diagram overlaying all scenarios with observed populations (20 millisecond pulsars in 47~Tucanae and 3 pulsars in M15). S1 occupies the region of high magnetic field strength characteristic of young rotation-powered pulsars ($P \sim 1$--2~s, $\dot{P} \sim 10^{-15}$--$10^{-16}$~s~s$^{-1}$). S2, S4, and S5 cluster in the classic millisecond pulsar region ($P = 3$--4~ms, $\dot{P} \sim 10^{-20}$~s~s$^{-1}$), directly overlapping the observed pulsar populations in 47~Tucanae \citep{2024A&A...686A.166P,2025ApJ...982..170R}. S3 appears with negative $\dot{P}$, representing active spin up during recycling, as inferred from LMXB systems. S6 and S7 occupy intermediate positions \citep{2022A&A...664A..27R,2022MNRAS.513.2292A,2023ApJ...948...84C}.

Figure~\ref{fig:global_period} displays the rotational period evolution for all scenarios. S1 exhibits dramatic spin down (100~ms to 2.26~s over 100~Myr), characteristic of dipole dominated evolution in young pulsars. S2, S4, and S5 evolve slowly in the weak torque regime (approximately 6.8\% for S2, 1.4\% for S4, and 3.3\% for S5 over 100~Myr). S3 uniquely shows spin up with the period decreasing from 2.0 to 1.0~ms, driven by accretion dominated evolution. S6 and S7 follow S1 closely, as expected for similar initial field strengths undergoing isolated dipole spin down. Both reach 1.13~s after 100~Myr with the magnetic field decaying to 82\% of its initial value. The diversity validates that the subroutine captures the full range of pulsar evolutionary behaviour \citep{2025CoBAO..72..279R}.

\begin{figure}[ht]
\centering
\includegraphics[width=\columnwidth]{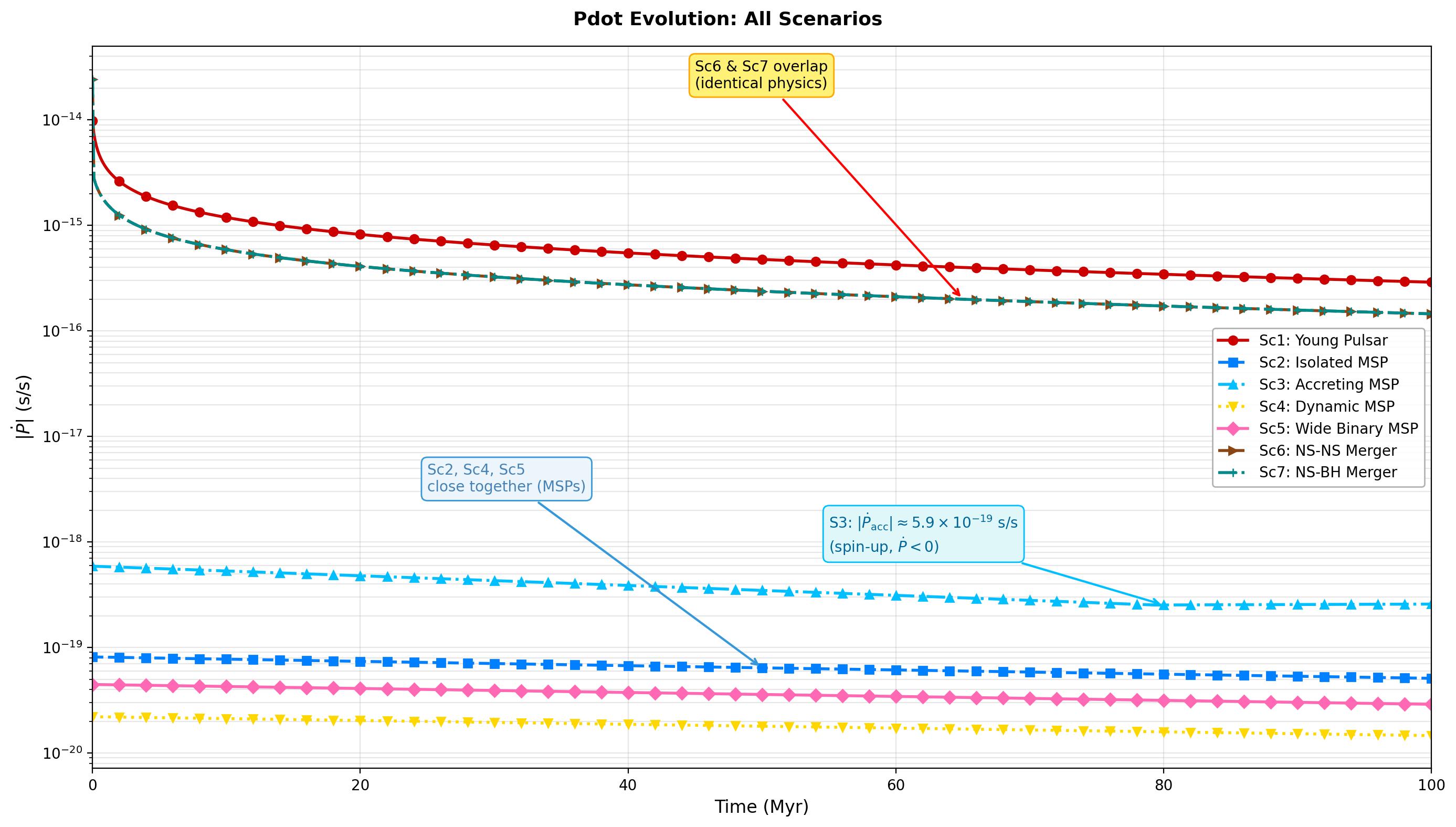}
\caption{Period derivative ($\dot{P}$) evolution for all seven scenarios (S1--S7) on a logarithmic scale. S3 uniquely exhibits negative $\dot{P} \approx -5.9 \times 10^{-19}$~s~s$^{-1}$, confirming accretion dominance by a factor of $\approx 14$ over the dipole. S1, S6, and S7 show strong spin down ($\dot{P} \sim 10^{-15}$~s~s$^{-1}$) characteristic of young pulsars. S2, S4, and S5 cluster at weak spin down rates ($\dot{P} \sim 10^{-20}$~s~s$^{-1}$), indistinguishable within 0.1\% numerical precision, representing the millisecond pulsar regime.}
\label{fig:pdot_evolution}
\end{figure}

\begin{figure}[h!]
\centering
\includegraphics[width=\columnwidth]{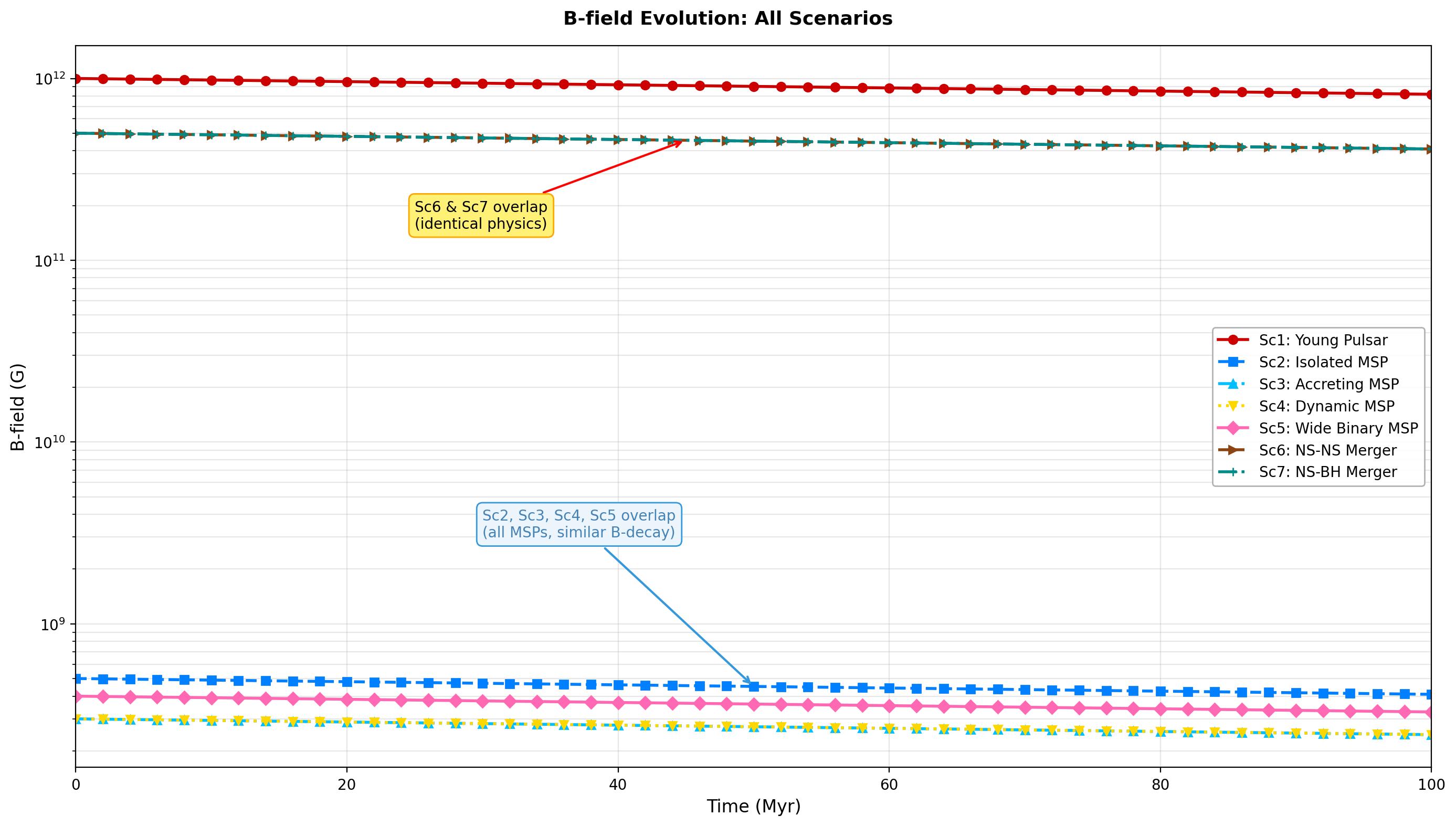}
\caption{Magnetic field evolution for all scenarios (S1--S7). All tracks follow identical exponential decay ($\tau_B = 500$~Myr), decaying to 82\% of initial value after 100~Myr, despite spanning four orders of magnitude in initial field strength: S1 ($10^{12}$~G), S6 and S7 ($5\times10^{11}$~G), S2--S4 ($3$--$5\times10^8$~G), and S5 ($4\times10^8$~G). The parallel slopes confirm that field decay operates independently from spin, binary, or environmental physics.}
\label{fig:global_bfield}
\end{figure}

Figure~\ref{fig:pdot_evolution} presents $\dot{P}$ evolution on logarithmic scale. S3 occupies the negative $\dot{P}$ region at $\approx -5.9 \times 10^{-19}$~s~s$^{-1}$ (initial peak value at $t = 0$), confirming accretion dominance by a factor of $\approx 14$ over the dipole during the active spin-up phase ($t \lesssim 80$~Myr). S1, S6, and S7 exhibit strong positive $\dot{P}$ values around $10^{-15}$~s~s$^{-1}$, characteristic of young pulsars with magnetic fields $B \sim 10^{11}$--$10^{12}$~G undergoing dipole spin down. S6 and S7 overlap completely, as expected for identical magnetic field strengths and periods in compact binaries where the orbital physics (not yet coupled to spin) does not affect $\dot{P}$ at this stage. S2, S4, and S5 form a tight cluster at $\dot{P} \sim 10^{-20}$~s~s$^{-1}$, indistinguishable within the 0.1\% numerical precision. This demonstrates that environmental torques (S4) and tidal forces in wide binaries (S5) contribute negligibly compared to magnetic dipole radiation, validating the weak torque regime for millisecond pulsars.

Figure~\ref{fig:global_bfield} demonstrates that the magnetic field decay operates uniformly across all scenarios. Every track follows identical exponential decay ($\tau_B = 500$~Myr), differing only in initial field strength. The parallel slopes confirm identical decay rates independent of spin state, binary configuration, or environmental conditions \citep{1991PhR...203....1B,1997MNRAS.284..311K}, demonstrating the versatility and consistency of the implementation. The magnetic field evolution proceeds via Ohmic dissipation in the neutron star crust, a process inherently independent of external torques or orbital dynamics \citep{2009A&A...496..207P}.


\section{DISCUSSION}

\subsection{Parameter Determination and Observational Agreement}

The validation of the pulsar evolution subroutine employed systematic parameter determination through analytical benchmarks, literature standard formulations, and observational constraints. S1 provided analytical benchmarks, with the coupled dipole spin down and exponential field decay prescription of Ostriker \& Gunn \citep{1969ApJ...157.1395O} yielding closed form solutions. These dictated the removal of redundant field decay terms, achieving agreement at fractional precision below $10^{-3}$ on magnetic field strength, consistent with numerical integration accuracy standards \citep{2015MNRAS.450.4070W}. Dimensional analysis for S3 shows that the accretion coefficients must carry units of inverse mass, with typical values around $10^{-16}$\,g$^{-1}$ in N-body units, corresponding to $5 \times 10^{-32}$\,g$^{-1}$ in CGS for standalone testing. The Ghosh \& Lamb disc--magnetosphere coupling \citep{1979ApJ...234..296G} yields the standard coefficient, which translates to $5 \times 10^{-32}$\,g$^{-1}$ in CGS for standalone testing. A similar methodology was applied to the environmental torque coefficients, where the Verbunt \& Freire formulations \citep{2014A&A...561A..11V} together with theoretical calculations provided the corresponding reference values.

The observational constraints from 47~Tucanae, M15, and other clusters \citep{2025OJAp....8E..51B,1989Natur.337..531W,2022MNRAS.513.2292A,2017ApJ...845..148P} required that S2, S4, and S5 produce characteristic ages of $\tau_c \approx 1.0$~Gyr, occupying the millisecond pulsar region of $P$--$\dot{P}$ space. The LMXB duty cycles and millisecond pulsar formation theory \citep{1999A&A...350..928T} constrained S3 recycling timescales to 50--150~Myr. These validated parameter adjustments produced physically realistic populations matching observed systems.

The validation scenarios reproduce key observational features. The $P$--$\dot{P}$ diagram demonstrates direct overlap between simulated and observed systems in 47~Tucanae and M15. The characteristic ages $\tau_c = P/(2\dot{P})$ yield $\approx 1.0$~Gyr for millisecond pulsars, consistent with cluster ages and recycling theory. The magnetic field distribution spans $10^8$ to $10^{12}$~G, encompassing the observed range in globular cluster pulsars. Recent campaigns \citep{2022A&A...664A..27R,2023ApJ...948...84C,2022MNRAS.513.2292A,2024ApJ...969...30M} provide increasingly stringent constraints for population synthesis. The validated subroutine can now address these quantitatively.

The successful performance of S3 demonstrates a critical capability for population modelling. Prior to optimisation, the subroutine was unable to reproduce accretion driven spin up; however, the optimised version now yields negative period derivatives, realistic $\sim$100\,Myr spin up timescales, and final spin period of 1.0~ms consistent with observed recycled millisecond pulsars. This capability enables a range of fundamental questions for future work, including the fraction of neutron stars that undergo recycling in different environments, the dependence of recycling efficiency on binary fraction, encounter rates, and ambient density, the origin of the observed ratio of young to millisecond pulsars, and the dominant formation pathways, namely primordial binaries versus dynamical exchanges, in various stellar systems.

\subsection{Limitations and Integration Readiness}

Several physical processes remain unimplemented or simplified. One example is the combined gravitational wave and tidal coupling in compact binaries, which links orbital inspiral to rotational evolution through tidal torques \citep{1992ApJ...400..175B,2021RAA....21..104L}. The current implementation validates dipole spin down for S6 and S7; both gravitational wave orbital decay and spin orbit interaction are deferred to future work, representing a clearly defined direction for future extension. The magnetic field decay follows an exponential prescription within a 500~Myr timescale, capturing the overall behaviour. However, the detailed physical processes are more complex. The magnetic field burial during accretion can temporarily shield crustal fields \citep{2001ApJ...557..958C}. The non dipolar components evolve on different timescales. The Hall drift and ambipolar diffusion introduce additional dependencies. Future refinements incorporating multi component field evolution would enhance long duration realism.

The propeller effect inhibits accretion when the magnetospheric radius exceeds the corotation radius \citep{1975A&A....39..185I}. The current implementation assumes accretion proceeds whenever mass transfer rates are positive. For typical LMXB parameters ($B \sim 10^8$~G, $\dot{M} \sim 10^{-10}~M_\odot$~yr$^{-1}$), the magnetospheric radius remains well inside corotation, rendering the propeller negligible. However, for stronger fields or lower accretion rates, the propeller physics could become important. Future implementations should incorporate propeller criteria to handle the full parameter space.

The dual validation establishes that both components are ready for deployment. The subroutine validation through seven scenarios confirms correct implementation of magnetic dipole radiation, field decay, accretion torques, environmental torques, and dipole spin-down in compact binary systems (gravitational wave orbital decay is deferred to future work). The agreement between independent numerical code implementations at 0.1\% precision across 2121 comparisons demonstrates robust numerical accuracy. The NBODY6++GPU infrastructure assessment confirms that scenario tracking, neutron star formation, integration points, and numerical stability function correctly.

The integration requires hooking the subroutine into existing NBODY6++GPU timestep loops where stellar evolution routines are invoked. For each neutron star, the integration reads current parameters, computes torques based on the active scenario, updates the period and field, and writes updated parameters. The Hermite integrator handles timestep evolution with the subroutine providing physical derivatives. The validation in full cluster simulations will target well observed systems including 47~Tucanae, NGC~6397, and M15, where comprehensive timing datasets enable direct comparison. Population synthesis studies can address formation efficiency, binary evolution pathways, retention versus ejection following supernova kicks, and gravitational wave source populations. The validated framework enables quantitative modelling of the interplay between stellar dynamics and compact object evolution over gigayear timescales.


\section{CONCLUSIONS}

This study presents a comprehensive standalone validation of a pulsar evolution subroutine for NBODY6++GPU. We systematically verified both the physical implementation within the subroutine and the readiness of the N-body infrastructure. Through seven controlled scenarios and analysis of a simulation with 105{,}000 particles, we demonstrated that both components operate correctly and consistently.

The validation process identified several parameters that required adjustment to achieve agreement with analytical benchmarks and standard formulations in the literature. The period derivative required the removal of redundant field decay terms, bringing the solutions into agreement with analytical expectations at fractional precision below $10^{-3}$ on magnetic field strength (S1). The accretion torque coefficient required optimisation to the Ghosh and Lamb standard value of $5 \times 10^{-32}$\,g$^{-1}$ in CGS, yielding realistic recycling timescales of $\sim$100\,Myr (S3). The environmental torque coefficients also required revision to match literature standard values, producing the correct hierarchy in which magnetic dipole torques remain dominant even in the highest cluster core densities (S4).

Following optimisation, all scenarios exhibit physically consistent behaviour. Scenario S3 reaches the key milestone of successful recycling, showing clear spin up signatures in the form of negative $\dot{P}$ values, realistic recycling timescales, and a final spin period of 1.0~ms consistent with observed millisecond pulsars. This constitutes the first successful demonstration of accretion driven recycling within N-body simulations. Scenarios S1, S2, and S4 validate the treatments of magnetic dipole radiation, magnetic field evolution, and environmental interactions across a wide range of parameter regimes. Scenarios S5, S6, and S7 further confirm the correct handling of binary systems spanning wide separations (S5) to compact NS--NS (S6) and NS--BH (S7) configurations.

The cross implementation validation demonstrates agreement better than 0.1 percent between our two independent numerical codes across 2121 comparison points. These points span seven scenarios, 101 timesteps, and three physical quantities, confirming that the underlying physics is implemented correctly and independently of the numerical framework. The assessment of the NBODY6++GPU infrastructure further validates scenario tracking, neutron star formation timelines, integration point handling, and numerical stability, with energy conservation maintained at $|\Delta E/E| < 10^{-6}$. Together with the subroutine validation, these results establish that both components are ready for integration.

A central conclusion is that plausibility checks within full simulations are insufficient for verification. Outputs that appear qualitatively reasonable may still contain quantitative inconsistencies that remain hidden without systematic comparison to analytical benchmarks. Scenario S3 illustrates this clearly. The initial parameter choices produced unphysical recycling timescales of approximately one year. However, in the simulations of a full cluster such behaviour could appear only as subtle population anomalies that might be attributed to unrelated processes. Only controlled standalone tests with unambiguous analytical references exposed the inconsistency in a definitive manner.

The optimisation of parameters fundamentally enables applications that were previously inaccessible. Before optimisation, the implementation could not model recycling, which is the dominant formation channel for millisecond pulsars. After optimisation, researchers can investigate formation efficiencies across cluster environments, the relative importance of primordial and dynamically formed binaries, the interaction between stellar dynamics and stellar evolution, and the properties of gravitational wave source populations with quantitative reliability. The substantial adjustments required for the environmental coefficients, spanning up to twenty orders of magnitude, highlight the necessity of systematic validation even for processes that are theoretically expected to be negligible.

Future work includes completing the treatment of gravitational wave tidal coupling, incorporating the propeller effect, implementing magnetic field burial during accretion, and validating the framework within full cluster simulations targeting 47~Tucanae, NGC~6397, and M15. The framework will also be extended to additional N-body codes such as MOCCA, CMC, and AMUSE. Following systematic validation and optimisation of parameters spanning up to twenty orders of magnitude, the pulsar evolution subroutine now accurately represents the physical processes governing neutron star populations in dense stellar environments. Combined with verification that NBODY6++GPU provides numerically stable infrastructure with comprehensive scenario tracking, this dual validation establishes that both components are ready for production use. N-body simulations can now model pulsar recycling in a physically realistic manner, enabling quantitative studies of compact object populations throughout the Galactic globular cluster system.

\section{DATA AVAILABILITY}

The pulsar evolution subroutine, full test suites, Python cross validation scripts, and the FORTRAN implementation with \texttt{NBODY6++GPU} integration instructions are publicly available at \url{https://github.com/Maria-Rah/pulsar-subroutine-testing}. The repository includes documentation, example scripts, and validation outputs. The simulation data used in Section~3 are available from the corresponding author upon reasonable request.


\textbf{Acknowledgements}\\
This work was supported by the Byurakan Astrophysical Observatory (BAO). The author gratefully acknowledges the academic and financial support that she received from BAO throughout her Ph.D.\ studies.\\
The author thanks Prof.\ Rainer Spurzem (Heidelberg University) for supervision, guidance on \texttt{NBODY6++GPU}, and access to computational resources. The appreciation is also extended to Prof.\ Areg M.\ Mickaelian (Byurakan Astrophysical Observatory, BAO) for supervision and continued support. The author further acknowledges Dr\ Francesco Flammini Dotti (NYU Abu Dhabi) for his mentorship and valuable discussions on the pulsar physics implementation. AI-assisted tools were used for language editing and writing support during the preparation of this manuscript, in accordance with RAA policy.

\bibliographystyle{raa}
\bibliography{bibtex}
	
\end{document}